\documentclass[aps,prr,reprint]{revtex4-2}

\usepackage{graphicx}
\usepackage{svg}
\usepackage{dcolumn}
\usepackage{bm}
\usepackage{braket}
\usepackage{amsmath}
\usepackage{amssymb}
\usepackage{hyperref}
\usepackage{url}
\usepackage[normalem]{ulem}

\hypersetup{hypertex=true,
colorlinks=true,
linkcolor=blue,
urlcolor = blue,
citecolor=blue}

\usepackage{color}

\begin{document}
\let\origsaddcontentsline\addcontentsline
\renewcommand{\addcontentsline}[3]{}

\title{Walsh-Floquet Theory of Periodic Kick Drives}

\author{James Walkling}
\email{jamwalk@pks.mpg.de}
\author{Marin Bukov}%
\affiliation{%
 Max Planck Institute for the Physics of Complex Systems, Nöthnitzer Strasse 38, 01187 Dresden, Germany
}%


\begin{abstract}
Periodic kick drives are ubiquitous in digital quantum control, computation, and simulation, and are instrumental in studies of chaos and thermalization for their efficient representation through discrete gates. However, in the commonly used Fourier basis, kick drives lead to poor convergence of physical quantities.
Instead, here we use the Walsh basis of periodic square-wave functions to describe the physics of periodic kick drives.  
In the strongly kicked regime, we find that it recovers Floquet dynamics of single- and many-body systems more accurately than the Fourier basis, due to the shape of the system's response in time. 
To understand this behavior, we derive an extended Sambe space formulation and an inverse-frequency expansion in the Walsh basis. 
We explain the enhanced performance within the framework of single-particle localization on the frequency lattice, where localization is correlated with small truncation errors.
We show that strong hybridization between states of the kicked system and Walsh modes gives rise to Walsh polaritons that can be studied on digital quantum simulators.  
Our work lays the foundations of Walsh-Floquet theory, which is naturally implementable on digital quantum devices and suited to Floquet state manipulation using discrete gates.
\end{abstract}

\maketitle

Quantum simulation is a pillar of quantum technology with numerous successes in the study of localization, thermalization, chaos, and topology. These simulations fall under two main categories: analogue and digital. For each of these categories, periodic driving is instrumental in the investigation of non-equilibrium physics within the paradigm of Floquet theory and the design of synthetic matter via ``Floquet engineering" \cite{Floquet1883, Bukov2015, Oka2019}. Analogue harmonic drives in optical lattices allow the simulation of gauge fields and topological models that are inaccessible in natural materials alongside dynamically induced localization \cite{Eckardt2017, BlochRev2017, AidelsburgerRev2018, SchaferRev2020}. On digital platforms, inherently non-equilibrium Floquet phases of matter can be simulated, such as discrete time crystals \cite{Sacha2017,Sondhi2019,Else2020, Yao2023,moon2024experimental}. Beside exactly implementing discrete stepwise-continuous protocols, arbitrary time dependence can be realized on digital platforms through Trotterization, which introduces an associated discretization (Trotter) error; it can cause Floquet heating in the system, and its mitigation or potential beneficial use is an active area of research \cite{Heyl2019, Wurtz2022, Kivlichan2020, Hongzheng2023}.

\begin{figure}[t!]
    \centering
    \includegraphics[scale=1]{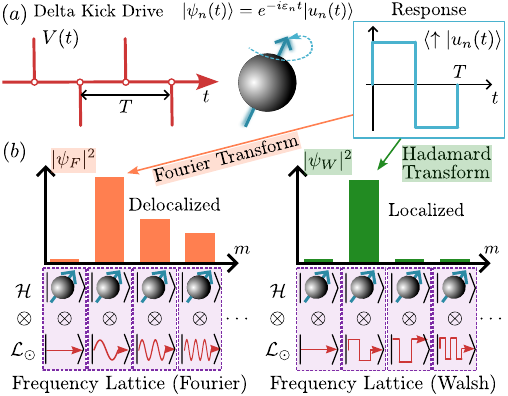}
    \caption{(a) A kicked system (represented as a spin-$1/2$) obeying the Schrödinger equation generically has a square wave response in the observables, via the micromotion $\ket{u_n(t)}$. (b) frequency lattice states in the tensor product of the physical spin Hilbert space, $\mathcal{H}$, and the space of periodic functions, $\mathcal{L}_\odot$ [schematic]. Depending on the choice of basis, the response is localized on the frequency lattice labeled by mode number, $m$. Wavefunctions are defined by $\braket{\uparrow |u_n(t)} = \sum \psi_f(m) f_m(t)$ where $f_m(t)$ represents Fourier or Walsh modes.
    }
    \label{fig:IntroFig}
\end{figure}

Periodic kick drives composed of delta function trains are exactly expressible on digital platforms. A paradigmatic example is the kicked Ising model, which is key in the simulation of exactly-solvable unitary dynamics \cite{Akila2016, Bertini2018, Bertini2019,Gopalakrishnan2019, AltmanRev2021, Ho2022}. Even beyond digital platforms, kick drives exhibit rich quantum dynamics; they give rise to non-ergodic and chaotic behavior~\cite{Haake1987, Chaudhury2009, Neill2016, Sieberer2019} and play a key role in large classes of quantum control protocols~\cite{Hahn1950,Tanner1968, Hans2020, Lasek2023}.

\begin{figure}[t]
    \centering
    \includegraphics[scale=1]{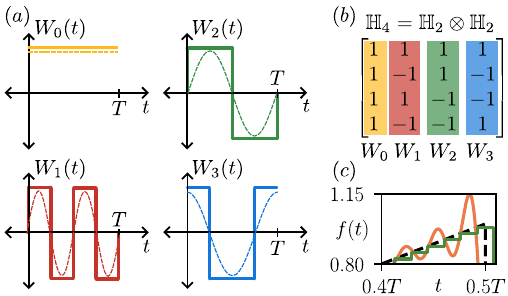}
    \caption{
    (a) Walsh basis functions for $N=4$ basis elements along with related trigonometric functions in a dashed line. This correspondence only works for $N=4$ since the roots of the Walsh functions are not evenly spaced for $N > 4$. 
    (b) construction of the Walsh basis from the Hadamard matrix for $N=4$. 
    (c) example of representing a function with discontinuities in different bases. The Walsh (green) consistently undershoots for smooth variation; by contrast, the Fourier  (orange) oscillates wildly near sharp discontinuities (Gibbs phenomenon) for a periodic sawtooth wave (dashed black).}
    \label{fig:Walsh}
\end{figure}

To solve perturbatively for the Floquet dynamics of such systems, a periodic basis with period $T=2\pi/\omega$ and frequency $\omega$ is used. In the frequency lattice construction \cite{Sambe1973,Eckardt2015}, the periodic part of the Floquet response -- the micromotion -- is expanded in the space of periodic functions $\mathcal{L}_\odot$, see~Fig.~\ref{fig:IntroFig}. The ``lattice" here refers to the labelling of Fourier basis functions $\exp(im\omega t)$ with $m \in \mathbb{Z}$. The frequency lattice is infinite; to study it numerically, the effective Hamiltonian matrix must be truncated, yielding basis-dependent results. For strongly periodically kicked systems, the commonly used Fourier series has notable disadvantages: a kick drive has constant Fourier series coefficients with no sensible cut-off; moreover, the cut-off process induces a strong ringing distortion called the ``Gibbs phenomenon" (see~Fig.~\ref{fig:Walsh}(c)) \cite{Hewitt1979}. While these effects can be mitigated, they point to the inadequacy of the Fourier basis to analyze digital drives.

In this work, we introduce the periodic Walsh functions $\{W_m(t)\}$ as an alternative basis for $\mathcal{L}_\odot$. The Walsh basis is a set of periodic, orthonormal, piecewise-constant functions that only take values $\pm 1$ as shown in Fig.~\ref{fig:Walsh}(a). Historically, it was used extensively in signal processing and has also seen a recent revival in physics~\cite{Walsh1923, Fine1949, golubov2012, Hayes2011, Shukla2023, zylberman2024, Pagano2024, Votto2024}. The Walsh basis truncates via discretizing the signal in real space with a time step $T/N$ where $N$ is the number of basis functions. Dynamical equations contain the time derivative; when treated explicitly as the time translation generator, we show that it takes a block diagonal form in the Walsh basis with an identical spectrum to the Fourier.

Under strong digital driving, quantum systems have a piecewise continuous response. We demonstrate that for a system driven with a delta function train, the response can be captured by far fewer non-zero coefficients in the Walsh basis as compared to the Fourier basis. Directly tied to this, we find that the error in the quasienergies can be orders of magnitude smaller in the Walsh than the Fourier basis when truncating to the same number of modes. To illustrate this, we calculate the time evolution of periodically-kicked single-particle and many-body mixed-field Ising model (MFIM).

We interpret the improved performance in the Walsh basis via a single-particle localization problem; for an accurate description of the dynamics under truncation of the frequency lattice, we find that the response must be localized in the frequency lattice to minimize the truncation error. We quantify localization through the participation entropy and demonstrate that strong localization on the frequency lattice corresponds to a more accurate description of the dynamics via the error in the quasienergies. 

Our work demonstrates that the Walsh basis can offer a marked advantage over the Fourier basis in the setting of strongly kicked systems. This has applications for digital quantum simulators where many toy models for chaos and thermalization use kick drives.

\textit{Walsh Basis in a Nutshell---}%
The Walsh basis $\{W_m(t)\}$ is a set of $N=2^n$ orthonormal step functions on $[0,T]$. For $n=1$, the Walsh basis elements are just the rows of the Hadamard matrix, 
\begin{equation}
\mathbb{H}_2 = \begin{bmatrix}
    1 & 1 \\
    1 & -1 \\
\end{bmatrix}.
\label{eqn:HadamardMatrix}
\end{equation}
To find the Walsh functions for larger $n$, one uses the Hadamard matrix of order $2n$ which is formed by taking the tensor product $n$ times of $\mathbb{H}_2$. An example of the corresponding matrix for $n=2$ is shown in Fig.~\ref{fig:Walsh}(b). These periodic functions are defined by a set of points $t_j = j T/2^n$ where $j=0,1,\cdots, 2^n-1$. Away from these points, the values of the Walsh functions come from interpolating in a piecewise constant manner (see Fig.~\ref{fig:Walsh}(a)). 
We label the states according to natural ordering \cite{Zhihua1983}; this means that the same function will change its label in the basis, for different $N$. For instance, we can label the Walsh function with a single root (e.g., $W_2(t)$ in Fig.~\ref{fig:Walsh}(a)) as $W_{N/2}(t)$, since this gives the correct label for any $N$. 
In Fig.~\ref{fig:Walsh}(a) and (b), we illustrate the Walsh functions on the real line and the character table from which they are derived for $n=2$. 

Both the Walsh and Fourier are complete bases such that their representations of functions should be equivalent; however, under truncation, the corresponding approximations differ. To render the frequency space finite, the Fourier basis applies a sharp cutoff which discards any coefficients associated with higher frequencies. By contrast, the Walsh basis maps different frequencies onto each other; the time discretization in the Walsh basis leads to distortion of the signal (aliasing) because it cannot resolve the shorter lengthscale features of a function \cite{Nyquist1928}. If higher frequencies carry significant weight, as in the presence of singularities, discarding higher frequencies in the Fourier basis suggests that the Walsh basis can outperform. Further details of the construction of the Walsh basis, and the difference in the Walsh and Fourier coefficients due to the aliasing effect are discussed in detail in the Supplementary Material. We now detail the use of the Walsh basis to solve the Floquet problem.

\textit{Quasienergy operator in the Walsh basis---}%
The solution to the Floquet problem, the evolution of a periodically driven system with Hamiltonian $\hat{H}(t) = \hat{H}(t+T)$, is given by the eigenstates and eigenvalues of the Floquet Hamiltonian, $H_F$, alongside the kick operator, $K(t)$, cf.~Supplementary Material. 

The eigenstates and eigenvalues of the Floquet Hamiltonian can be found from the quasienergy operator,
\begin{equation}
    \hat{Q}(t) = \hat{H}(t) - i\partial_t.
    \label{eqn:Qmatrix}
\end{equation}
To diagonalize it, we write it as $\bar{Q}$ in an extended frequency space (Sambe space \cite{Sambe1973}), $\mathcal{F}= \mathcal{H} \otimes \mathcal{L}_\odot$. $\mathcal{H}$ is the physical Hilbert space associated with $\hat{H}(t)$, and we promote the periodic time-dependence to its own Hilbert-space, $\mathcal{L}_\odot$ \cite{Eckardt2015}. The inner product on $\mathcal{F}$, is $\langle \langle u | v \rangle \rangle = T^{-1}\int_0^T \text{d} t \braket{v(t)|u(t)}$.

The operator $\bar{Q}$ has eigenvalues $\varepsilon_{nm} = \varepsilon_n + m\omega$, where $m \in \mathbb{Z}$. These eigenvalues, known as quasienergies, are defined modulo $m\omega$. In figures, we rescale to define dimensionless Floquet phases $\theta_n=\varepsilon_n T$. The corresponding eigenvectors, the Floquet modes, are expressed as $\ket{u_{nm}(t)} = e^{im\omega t} \ket{u_n(t)}$, with the phase factor accounting for the modularity. The integer $m$ is referred to as a photon index, since it can be intuitively thought of as the number of photons dressing the state. Previously, the Fourier basis has been used to decompose $\ket{u(t)}$ as $\ket{\alpha m(t)}= \ket{\alpha}e^{im\omega t}$, and these are expressed as $\ket{\alpha,m}\rangle$ in the extended space.

To instead write the operator $\bar{Q}$ in the Walsh basis, one needs to handle the derivative operator with care; on a piecewise-constant basis, the derivative is not necessarily equal to the generator of continuous time translations as it is for a continuous basis. Indeed, in the Walsh basis, the derivative is not the generator of translations for finite $N$, as becomes evident from their different spectra. However, the translation generator, $\hat{G}$, defined thoroughly in the Supplementary Material, agrees exactly with the truncated spectrum of the derivative for any $N$. Hence, in the Walsh basis, $\hat{G}$ is used in place of the derivative in Eq.~\eqref{eqn:Qmatrix}; in the extended space associated with the Walsh basis, Eq.~\ref{eqn:Qmatrix} reads as $\bar{Q}=H-i\hat{G}$. Properties of $\hat{G}$ in the Walsh basis, such as its block-diagonality, and truncation of the frequency lattice are discussed further in the Supplementary Material.

\begin{figure}[t!]
    \centering
    \includegraphics[scale=1]{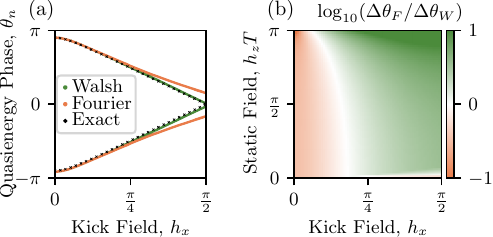}
    \caption{Single-particle quasienergies from a kick drive, see Fig.~\ref{fig:IntroFig}(a). 
    (a) is the quasienergy spectrum for $\omega =10$ and $h_z=4.5$ showing the superior performance of the Walsh basis over the Fourier. 
    (b) quantifies the error over a range of parameters, showing that Walsh outperforms Fourier the most for strong kick drives. In the orange region, the Fourier basis better approximates the quasienergies, and in the green region the Walsh performs better. $N=32(31)$ modes used in both plots.
    }
    \label{fig:SPerror}
\end{figure}

\begin{figure}[t]
    \centering
    \includegraphics[scale=0.99]{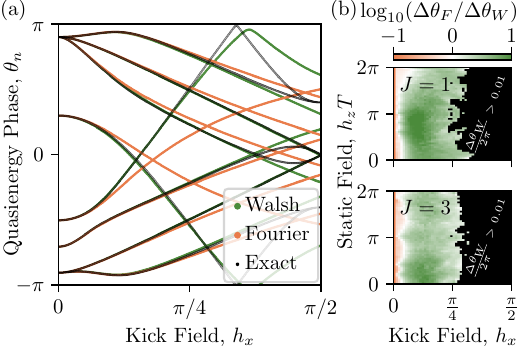}
    \caption{(a) Many-body quasienergy spectrum for $L=3$ spins and $N{=}32(31)$ modes with $J=1$ and $h_zT=1.1\pi$. The Walsh basis outperforms the Fourier. (b) Comparison of the error in the dimensionless quasienergy phases calculated using the Walsh and Fourier basis to the same level of truncation, $N{=}64(63)$ modes, for $L=6$ spins. Green indicates that Walsh is more accurate than the Fourier, and orange means the opposite.
    Apart from very small kick fields, the Walsh basis outperforms over a large parameter regime. At large kick fields, in the black region, there is an error w.r.t.~the exact solution by more than 1\% in both bases.
    }
    \label{fig:MBerror}
\end{figure}

\textit{Bases comparison for a kicked Ising model---}%
We focus on a kick drive which is an up-down kick with zero average, $V(t)$, as illustrated in red in Fig.~\ref{fig:IntroFig}(a); this choice is motivated by its anti-derivative being a single Walsh mode, and is made for simplicity. The $\bar{Q}$ operator for the Fourier and Walsh bases is shown in the Supplementary Material for the case of this kick drive.

To compare the bases, we quantify their accuracy in reproducing the Floquet dynamics via the quasienergy phases $\theta_n$ for the MFIM:
\begin{equation}
    H = - \sum_{i=0}^{L-2} J \sigma_i^z \sigma_{i+1}^z 
    - \sum_{i=0}^{L-1} \big( h_z \sigma_i^z + h_x V(t) \sigma_i^x \big).
    \label{eqn:MFIM}
\end{equation}
The error is defined by subtracting off the exact result: $\Delta \theta_\text{basis}=|\theta_{\text{basis}}-\theta_{\text{exact}}|$.
When comparing the Walsh with the Fourier basis Walsh always has an even number of modes (i.e., $N=32$), whereas the Fourier basis benefits from having an odd number due to symmetry (i.e., $N=31$).

First comparing the two in the single-particle limit of Eq.~\eqref{eqn:MFIM} ($J=0$ and $L=1$), we plot the quasienergy spectrum and errors at finite truncation in Fig.~\ref{fig:SPerror}(a) and (b) respectively. Although the Fourier basis outperforms the Walsh basis for weakly kicked drives, the Walsh performs much better than the Fourier basis for strongly kicked drives. For zero field strength, the response of the spin, $\ket{u(t)}$, is time-independent, and this is a single mode in both bases; hence, they perform equally well. For weak kicking, the smooth response is marginally better captured by the Fourier (Fig.~\ref{fig:SPerror}(a) for small kick field shows very little difference). When $h_z \sim \pi/4$, the discontinuities are significant enough to favor the Walsh basis. Along the line $h_z=0$, the exact quasienergies are zero which is perfectly calculated by Fourier due to its reflection symmetry for an odd number of modes.

In the many-body case (Fig.~\ref{fig:MBerror}(a)), we see that the Walsh basis (green) is orders of magnitude closer to the exact numerics than the Fourier (orange) for intermediate kick strengths less than $\pi/2$. In addition, the Walsh basis (green) outperforms the Fourier basis (orange) over a broad parameter range as illustrated in Fig.~\ref{fig:MBerror}(b). For larger values of the kick field, $h_x$~\footnote{$h_x$ is a dimensionless quantity in natural units since it is the amplitude of a delta function drive. $h_z$ is dimensionful since it just multiplies some time independent term.}, the Walsh basis outperforms by at least an order of magnitude. These larger $h_x$ are relevant to several models involving kick drives to explore thermalization, localization and non-equilibrium phases since this is where the drive plays a significant role. Alongside the qualitative reasons we discussed for the poor behavior of the Fourier basis in the presence of kick drives, we now give compelling reasons why the Walsh is the appropriate basis.

\textit{Understanding errors via Frequency lattice localization---}%
There are several intuitive explanations for why the Walsh basis outperforms the continuous Fourier basis for kick drives. Since kick drives are periodic in frequency space with constant Fourier coefficients $c_n \sim \mathcal{O}(1)$, imposing a cutoff is a poor approximation since we discard infinitely many relevant frequencies. The aliasing effect creates several copies of the true spectrum (see Supplementary Material), which preserves the translational symmetry of the kick drive in frequency space, giving the Walsh basis an advantage in representing these singular functions at finite truncation. 

Physically, accurate reproduction of the dynamics under truncation can be understood through localization on the frequency lattice, $\mathcal{L}_\odot$. The response of the system to the drive, encoded by the Floquet states $\ket{u_n(t)}$, can have many modes when expanded in a given time-periodic basis. Through truncation, we implement a cutoff on these modes. Hence, if $\ket{u_n(t)}$ are well-localized in photon space for a given basis, we expect to get a good approximation by truncating in that basis. We emphasize that the basis should be optimized to the response, $\ket{u_n(t)}$, rather than the drive. Rather surprisingly, this means that the Walsh basis is a poor basis to capture dynamics induced by square wave drives despite having a single mode spectrum (see Supplementary Material).

\begin{figure}[t!]
    \centering
    \includegraphics[width=\linewidth]{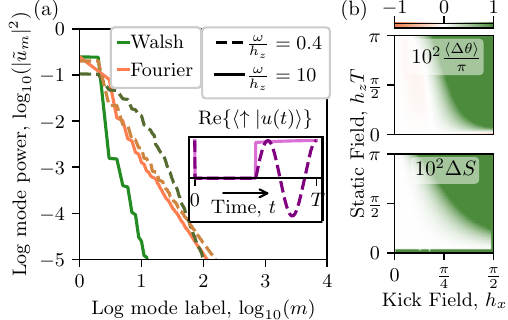}
    \caption{(a) Spin-up component of the response of a kicked two-level system system, defined via $\braket{\uparrow| u(t)} = \sum_m \tilde{u}_m f_m(t)$, where $f_m(t)$ denotes the orthonormal basis (Fourier, Walsh), see legend. Main plot shows the modes; the inset -- their time evolution, which is close to a square wave at high frequencies. Dashed (solid) lines indicate the low- (high-) frequency regime. At high frequency, the signal is strongly localized in the Walsh basis, and follows a $1/m$ power law decay otherwise. (b) difference in quasienergy errors, $\langle \Delta \theta \rangle \equiv\Delta \theta_{FW}=|\Delta \theta_F|- |\Delta \theta_W|$, and photon participation entropy ($\Delta S$ Eq.~\eqref{eqn:PPE}) highlighting the similarity between localization and associated error, scaled by a factor of $10^2$ (see text box). The errors are computed over all spins; localization is only shown for the spin-up component. The singular behavior near $h_zT=0$ comes from the degeneracy of $U(T,h_z=0)=\mathbf{1}.$
    Data obtained using Eq.~\eqref{eqn:MFIM} with $L=1$; for (a) $\omega, h_x= 10, \pi/2$.}
    \label{fig:localization}
\end{figure}

In the Fourier case, the localization problem is a general Wannier-Stark model in 1D; there is a linearly varying on-site potential (due to $-i\partial_t \equiv -i \hat{G}$) along with some tunneling elements that appear off-diagonal corresponding to an effective Hamiltonian in Sambe space $\bar{H}= \sum_j j \omega \mathbb{I} \otimes \hat{a}^\dagger_j \hat{a}_{j} + \sum_{j,j'} \hat{H}_{j-j'} \otimes \hat{a}^\dagger_j \hat{a}_{j'},$ where the operator $\hat{a}^\dagger_j$ creates a virtual ``photon" with frequency $j\omega$ on the space $\mathcal{L}_\odot$. 

The case of a harmonic drive is well-approximated by the Fourier basis; this corresponds to nearest neighbor tunnelling and exponentially localized eigenstates. By contrast, a generic kick drive has eigenstates that do not appear to decay with distance at all in the Fourier basis due to the all-to-all coupling in $\bar{H}$ with constant magnitude. 

For the Walsh basis, the non-diagonal form of the derivative means that the mapping to a Wannier-Stark ladder is no longer possible. However, in the high-frequency regime, it is possible to approximate the time-dependence of the response, $\ket{u_n(t)}= \exp(- i K(t)) \ket{u_n(0)}$, using the van Vleck expansion \cite{Bukov2015,Goldman2014}. The time evolution of the response is generated by the kick operator $K(t)$,
\begin{align}
    \exp(-i K(t)) = \exp \bigg[ -i\int^t \text{d}s \, H(s) \bigg] + \mathcal{O}(\omega^{-2}).
    \label{eqn:kickop}
\end{align}
Hence, in the high-frequency regime, using Eq.~\eqref{eqn:kickop}, we can guess at the form of the time-evolution and pick a basis accordingly. In particular, the Walsh basis performs well for the kick drive $V(t) =\dot{W}_{N/2}(t)$ since the response, $e^{-iK(t)} {\propto} W_{N/2}(t) + \mathcal{O}(\omega^{-2})$ (inset of Fig.~\ref{fig:localization}(a)). 
Away from the high-frequency regime, we can turn to other indicators of localization. 

We can view frequency lattice localization as a semi-quantitative predictor for the error in the quasienergies. 
The errors can be quantified via the localization of the modes, $\ket{u_n(t)}$, using the photon participation entropy,
\begin{equation}
    S= -\sum_{m} P_m \ln(P_m),
    \label{eqn:PPE}
\end{equation} 
where we trace over states $\ket{\alpha}\in\mathcal{H}$ to define a probability of occupying the $m$th mode, $P_m=\sum_\alpha|\langle \braket{\alpha m|\alpha m}\rangle|^2$, independent of the physical state $\ket{\alpha}$. While other localization measures such as the inverse participation ration (IPR) also capture this localization, the logarithm present in the photon participation entropy makes it drop off in value less steeply for extended states and leads to a marginally better comparison with the error.
In Fig.~\ref{fig:localization}(b) and (c), we see that a large photon participation entropy (delocalized in the up spin component) corresponds to a larger error. The participation entropy qualitatively predicts when solutions from truncation are more accurate. This demonstrates clearly how choosing the optimal basis of $\mathcal{L}_\odot$ for a given drive can be understood as a problem in single-particle localization physics. We have omitted the spin-down component since this suffers from resonance effects in the first Floquet Brillouin zone. 

Resonances in the spectrum can lead to disagreement between error and delocalization due to strong hybridization between physical and photonic degrees of freedom, a.k.a.~polariton states; averaging over the physical degrees of freedom becomes a coarse approximation. Depending on parameters, the polaritons behave differently. Arbitrarily weak drive generically gives a ``Fourier" polariton where one spin component hybridizes with $\exp(i m\omega t)$. However, for kick field $h_x=\pi/2$ and $h_z T\ll1$, one observes a ``Walsh" polariton where a component hybridizes with $W_m(t)$.

Both the spin-down component and polariton are discussed further in the Supplementary Material. 

\textit{Discussion and Outlook---}%
We have shown that Floquet analysis can be performed consistently in the Walsh basis. For kicked drives, we demonstrated that the Walsh basis more accurately describes the time evolution, when compared to Fourier, at a given truncation level in both single- and many-particle systems. The superior performance of the Walsh basis can be traced back to the response of the system (rather than the drive): the square-wave time-dependence of the Floquet modes is naturally expressed in the Walsh basis. Physically, we explain the increase in truncation error as a delocalization of the Floquet modes on the frequency lattice ($\mathcal{L}_\odot$); measures of localization, such as the photon participation entropy, capture the relative accuracy of periodic bases in describing the Floquet dynamics. 

While localization entropy generally predicts the error well, it can jump sharply near resonances. In the high-frequency regime, resonances introduce factors $e^{im\omega t}$ into the response, delocalizing it across several modes in non-Fourier bases, even if it remains confined within a particular Floquet zone. When states of $\mathcal{H}$ in a Floquet zone have different localization behavior, tracing over $\mathcal{H}$ can lead to an incorrect error estimate. Moreover, convergence is not generally guaranteed in the Fourier or the Walsh basis, as the number of modes $N\to\infty$. However, there is still asymptotic validity in the solutions obtained for finite $N$.

In practice, the Walsh basis has recently seen a surge of interest in several research areas of quantum physics. 
It is being actively investigated for the advantages it offers over alternative bases in both theoretical and experimental settings in quantum sensing and control~\cite{Wang2024, Pagano2024}. The Walsh basis offers a promising tool for designing digital shortcuts-to-adiabaticity on quantum simulators, such as by extending Floquet counterdiabatic driving~\cite{Schindler2024,schindler2024geometric, xichen2024, Sun2022, Wurtz2022} to periodically kicked drives.
Similarly, in quantum computation and simulation, recent works have investigated the Walsh basis to efficiently encode the action of Pauli gates~\cite{Votto2024, Georges2025, zylberman2024}. Our study gives rise to a natural reformulation of the inverse-frequency expansion, which reduces the computational cost of effective Hamiltonian calculations by eliminating contributions from certain expansion orders (see Supplementary Material). 
Curiously, interaction with the drive can also give rise to Walsh polaritons (Fig.~S1 (d)) in the presence of particularly strong kicks -- where spin or particle degrees of freedom hybridize with the drive in one of the Walsh modes; this opens up exciting and natural new possibilities for studying polariton physics on digital quantum simulators.

Advancing the toolbox of periodic drives, we have quantified the performance of the Walsh basis and established it as a competitive new tool. Our work opens up new pathways to think about and utilize digital Floquet drives on NISQ devices.

\begin{acknowledgments}
We thank A.~Eckardt, M.~Heyl, and R.~Moessner for fruitful discussions. 
Funded by the European Union (ERC, QuSimCtrl, 101113633). Views and opinions expressed are however those of the authors only and do not necessarily reflect those of the European Union or the European Research Council Executive Agency. Neither the European Union nor the granting authority can be held responsible for them.
Numerical simulations were performed on the MPIPKS HPC cluster.
\end{acknowledgments}


\bibliography{apssamp}

\begin{thebibliography}{57}%
\makeatletter
\providecommand \@ifxundefined [1]{%
 \@ifx{#1\undefined}
}%
\providecommand \@ifnum [1]{%
 \ifnum #1\expandafter \@firstoftwo
 \else \expandafter \@secondoftwo
 \fi
}%
\providecommand \@ifx [1]{%
 \ifx #1\expandafter \@firstoftwo
 \else \expandafter \@secondoftwo
 \fi
}%
\providecommand \natexlab [1]{#1}%
\providecommand \enquote  [1]{``#1''}%
\providecommand \bibnamefont  [1]{#1}%
\providecommand \bibfnamefont [1]{#1}%
\providecommand \citenamefont [1]{#1}%
\providecommand \href@noop [0]{\@secondoftwo}%
\providecommand \href [0]{\begingroup \@sanitize@url \@href}%
\providecommand \@href[1]{\@@startlink{#1}\@@href}%
\providecommand \@@href[1]{\endgroup#1\@@endlink}%
\providecommand \@sanitize@url [0]{\catcode `\\12\catcode `\$12\catcode `\&12\catcode `\#12\catcode `\^12\catcode `\_12\catcode `\%12\relax}%
\providecommand \@@startlink[1]{}%
\providecommand \@@endlink[0]{}%
\providecommand \url  [0]{\begingroup\@sanitize@url \@url }%
\providecommand \@url [1]{\endgroup\@href {#1}{\urlprefix }}%
\providecommand \urlprefix  [0]{URL }%
\providecommand \Eprint [0]{\href }%
\providecommand \doibase [0]{https://doi.org/}%
\providecommand \selectlanguage [0]{\@gobble}%
\providecommand \bibinfo  [0]{\@secondoftwo}%
\providecommand \bibfield  [0]{\@secondoftwo}%
\providecommand \translation [1]{[#1]}%
\providecommand \BibitemOpen [0]{}%
\providecommand \bibitemStop [0]{}%
\providecommand \bibitemNoStop [0]{.\EOS\space}%
\providecommand \EOS [0]{\spacefactor3000\relax}%
\providecommand \BibitemShut  [1]{\csname bibitem#1\endcsname}%
\let\auto@bib@innerbib\@empty
\bibitem [{\citenamefont {Floquet}(1883)}]{Floquet1883}%
  \BibitemOpen
  \bibfield  {author} {\bibinfo {author} {\bibfnamefont {G.}~\bibnamefont {Floquet}},\ }\bibfield  {title} {\bibinfo {title} {Sur les équations différentielles linéaires à coefficients périodiques},\ }\href {https://doi.org/10.24033/asens.220} {\bibfield  {journal} {\bibinfo  {journal} {Annales scientifiques de l’École normale supérieure}\ }\textbf {\bibinfo {volume} {12}},\ \bibinfo {pages} {47–88} (\bibinfo {year} {1883})}\BibitemShut {NoStop}%
\bibitem [{\citenamefont {Bukov}\ \emph {et~al.}(2015)\citenamefont {Bukov}, \citenamefont {D’Alessio},\ and\ \citenamefont {Polkovnikov}}]{Bukov2015}%
  \BibitemOpen
  \bibfield  {author} {\bibinfo {author} {\bibfnamefont {M.}~\bibnamefont {Bukov}}, \bibinfo {author} {\bibfnamefont {L.}~\bibnamefont {D’Alessio}},\ and\ \bibinfo {author} {\bibfnamefont {A.}~\bibnamefont {Polkovnikov}},\ }\bibfield  {title} {\bibinfo {title} {Universal high-frequency behavior of periodically driven systems: from dynamical stabilization to floquet engineering},\ }\href {https://doi.org/10.1080/00018732.2015.1055918} {\bibfield  {journal} {\bibinfo  {journal} {Advances in Physics}\ }\textbf {\bibinfo {volume} {64}},\ \bibinfo {pages} {139–226} (\bibinfo {year} {2015})}\BibitemShut {NoStop}%
\bibitem [{\citenamefont {Oka}\ and\ \citenamefont {Kitamura}(2019)}]{Oka2019}%
  \BibitemOpen
  \bibfield  {author} {\bibinfo {author} {\bibfnamefont {T.}~\bibnamefont {Oka}}\ and\ \bibinfo {author} {\bibfnamefont {S.}~\bibnamefont {Kitamura}},\ }\bibfield  {title} {\bibinfo {title} {Floquet engineering of quantum materials},\ }\href {https://doi.org/10.1146/annurev-conmatphys-031218-013423} {\bibfield  {journal} {\bibinfo  {journal} {Annual Review of Condensed Matter Physics}\ }\textbf {\bibinfo {volume} {10}},\ \bibinfo {pages} {387–408} (\bibinfo {year} {2019})}\BibitemShut {NoStop}%
\bibitem [{\citenamefont {Eckardt}(2017)}]{Eckardt2017}%
  \BibitemOpen
  \bibfield  {author} {\bibinfo {author} {\bibfnamefont {A.}~\bibnamefont {Eckardt}},\ }\bibfield  {title} {\bibinfo {title} {Colloquium: Atomic quantum gases in periodically driven optical lattices},\ }\bibfield  {journal} {\bibinfo  {journal} {Reviews of Modern Physics}\ }\textbf {\bibinfo {volume} {89}},\ \href {https://doi.org/10.1103/revmodphys.89.011004} {10.1103/revmodphys.89.011004} (\bibinfo {year} {2017})\BibitemShut {NoStop}%
\bibitem [{\citenamefont {Gross}\ and\ \citenamefont {Bloch}(2017)}]{BlochRev2017}%
  \BibitemOpen
  \bibfield  {author} {\bibinfo {author} {\bibfnamefont {C.}~\bibnamefont {Gross}}\ and\ \bibinfo {author} {\bibfnamefont {I.}~\bibnamefont {Bloch}},\ }\bibfield  {title} {\bibinfo {title} {Quantum simulations with ultracold atoms in optical lattices},\ }\href@noop {} {\bibfield  {journal} {\bibinfo  {journal} {Science}\ }\textbf {\bibinfo {volume} {357}},\ \bibinfo {pages} {995} (\bibinfo {year} {2017})}\BibitemShut {NoStop}%
\bibitem [{\citenamefont {Aidelsburger}\ \emph {et~al.}(2018)\citenamefont {Aidelsburger}, \citenamefont {Nascimbene},\ and\ \citenamefont {Goldman}}]{AidelsburgerRev2018}%
  \BibitemOpen
  \bibfield  {author} {\bibinfo {author} {\bibfnamefont {M.}~\bibnamefont {Aidelsburger}}, \bibinfo {author} {\bibfnamefont {S.}~\bibnamefont {Nascimbene}},\ and\ \bibinfo {author} {\bibfnamefont {N.}~\bibnamefont {Goldman}},\ }\bibfield  {title} {\bibinfo {title} {Artificial gauge fields in materials and engineered systems},\ }\href {https://doi.org/https://doi.org/10.1016/j.crhy.2018.03.002} {\bibfield  {journal} {\bibinfo  {journal} {Comptes Rendus Physique}\ }\textbf {\bibinfo {volume} {19}},\ \bibinfo {pages} {394} (\bibinfo {year} {2018})},\ \bibinfo {note} {quantum simulation / Simulation quantique}\BibitemShut {NoStop}%
\bibitem [{\citenamefont {Sch{\"a}fer}\ \emph {et~al.}(2020)\citenamefont {Sch{\"a}fer}, \citenamefont {Fukuhara}, \citenamefont {Sugawa}, \citenamefont {Takasu},\ and\ \citenamefont {Takahashi}}]{SchaferRev2020}%
  \BibitemOpen
  \bibfield  {author} {\bibinfo {author} {\bibfnamefont {F.}~\bibnamefont {Sch{\"a}fer}}, \bibinfo {author} {\bibfnamefont {T.}~\bibnamefont {Fukuhara}}, \bibinfo {author} {\bibfnamefont {S.}~\bibnamefont {Sugawa}}, \bibinfo {author} {\bibfnamefont {Y.}~\bibnamefont {Takasu}},\ and\ \bibinfo {author} {\bibfnamefont {Y.}~\bibnamefont {Takahashi}},\ }\bibfield  {title} {\bibinfo {title} {Tools for quantum simulation with ultracold atoms in optical lattices},\ }\href@noop {} {\bibfield  {journal} {\bibinfo  {journal} {Nat. Rev. Phys.}\ }\textbf {\bibinfo {volume} {2}},\ \bibinfo {pages} {411} (\bibinfo {year} {2020})}\BibitemShut {NoStop}%
\bibitem [{\citenamefont {Sacha}\ and\ \citenamefont {Zakrzewski}(2017)}]{Sacha2017}%
  \BibitemOpen
  \bibfield  {author} {\bibinfo {author} {\bibfnamefont {K.}~\bibnamefont {Sacha}}\ and\ \bibinfo {author} {\bibfnamefont {J.}~\bibnamefont {Zakrzewski}},\ }\bibfield  {title} {\bibinfo {title} {Time crystals: a review},\ }\href {https://doi.org/10.1088/1361-6633/aa8b38} {\bibfield  {journal} {\bibinfo  {journal} {Reports on Progress in Physics}\ }\textbf {\bibinfo {volume} {81}},\ \bibinfo {pages} {016401} (\bibinfo {year} {2017})}\BibitemShut {NoStop}%
\bibitem [{\citenamefont {Khemani}\ \emph {et~al.}(2019)\citenamefont {Khemani}, \citenamefont {Moessner},\ and\ \citenamefont {Sondhi}}]{Sondhi2019}%
  \BibitemOpen
  \bibfield  {author} {\bibinfo {author} {\bibfnamefont {V.}~\bibnamefont {Khemani}}, \bibinfo {author} {\bibfnamefont {R.}~\bibnamefont {Moessner}},\ and\ \bibinfo {author} {\bibfnamefont {S.~L.}\ \bibnamefont {Sondhi}},\ }\href {https://doi.org/10.48550/ARXIV.1910.10745} {\bibinfo {title} {A brief history of time crystals}} (\bibinfo {year} {2019})\BibitemShut {NoStop}%
\bibitem [{\citenamefont {Else}\ \emph {et~al.}(2020)\citenamefont {Else}, \citenamefont {Monroe}, \citenamefont {Nayak},\ and\ \citenamefont {Yao}}]{Else2020}%
  \BibitemOpen
  \bibfield  {author} {\bibinfo {author} {\bibfnamefont {D.~V.}\ \bibnamefont {Else}}, \bibinfo {author} {\bibfnamefont {C.}~\bibnamefont {Monroe}}, \bibinfo {author} {\bibfnamefont {C.}~\bibnamefont {Nayak}},\ and\ \bibinfo {author} {\bibfnamefont {N.~Y.}\ \bibnamefont {Yao}},\ }\bibfield  {title} {\bibinfo {title} {Discrete time crystals},\ }\href {https://doi.org/10.1146/annurev-conmatphys-031119-050658} {\bibfield  {journal} {\bibinfo  {journal} {Annual Review of Condensed Matter Physics}\ }\textbf {\bibinfo {volume} {11}},\ \bibinfo {pages} {467–499} (\bibinfo {year} {2020})}\BibitemShut {NoStop}%
\bibitem [{\citenamefont {Zaletel}\ \emph {et~al.}(2023)\citenamefont {Zaletel}, \citenamefont {Lukin}, \citenamefont {Monroe}, \citenamefont {Nayak}, \citenamefont {Wilczek},\ and\ \citenamefont {Yao}}]{Yao2023}%
  \BibitemOpen
  \bibfield  {author} {\bibinfo {author} {\bibfnamefont {M.~P.}\ \bibnamefont {Zaletel}}, \bibinfo {author} {\bibfnamefont {M.}~\bibnamefont {Lukin}}, \bibinfo {author} {\bibfnamefont {C.}~\bibnamefont {Monroe}}, \bibinfo {author} {\bibfnamefont {C.}~\bibnamefont {Nayak}}, \bibinfo {author} {\bibfnamefont {F.}~\bibnamefont {Wilczek}},\ and\ \bibinfo {author} {\bibfnamefont {N.~Y.}\ \bibnamefont {Yao}},\ }\bibfield  {title} {\bibinfo {title} {Colloquium: Quantum and classical discrete time crystals},\ }\href {https://doi.org/10.1103/RevModPhys.95.031001} {\bibfield  {journal} {\bibinfo  {journal} {Rev. Mod. Phys.}\ }\textbf {\bibinfo {volume} {95}},\ \bibinfo {pages} {031001} (\bibinfo {year} {2023})}\BibitemShut {NoStop}%
\bibitem [{\citenamefont {Moon}\ \emph {et~al.}(2024)\citenamefont {Moon}, \citenamefont {Schindler}, \citenamefont {Sun}, \citenamefont {Druga}, \citenamefont {Knolle}, \citenamefont {Moessner}, \citenamefont {Zhao}, \citenamefont {Bukov},\ and\ \citenamefont {Ajoy}}]{moon2024experimental}%
  \BibitemOpen
  \bibfield  {author} {\bibinfo {author} {\bibfnamefont {L.~J.~I.}\ \bibnamefont {Moon}}, \bibinfo {author} {\bibfnamefont {P.~M.}\ \bibnamefont {Schindler}}, \bibinfo {author} {\bibfnamefont {Y.}~\bibnamefont {Sun}}, \bibinfo {author} {\bibfnamefont {E.}~\bibnamefont {Druga}}, \bibinfo {author} {\bibfnamefont {J.}~\bibnamefont {Knolle}}, \bibinfo {author} {\bibfnamefont {R.}~\bibnamefont {Moessner}}, \bibinfo {author} {\bibfnamefont {H.}~\bibnamefont {Zhao}}, \bibinfo {author} {\bibfnamefont {M.}~\bibnamefont {Bukov}},\ and\ \bibinfo {author} {\bibfnamefont {A.}~\bibnamefont {Ajoy}},\ }\bibfield  {title} {\bibinfo {title} {Experimental observation of a time rondeau crystal: Temporal disorder in spatiotemporal order},\ }\href {https://arxiv.org/abs/2404.05620} {\bibfield  {journal} {\bibinfo  {journal} {arXiv preprint arXiv:2404.05620}\ } (\bibinfo {year} {2024})}\BibitemShut {NoStop}%
\bibitem [{\citenamefont {Heyl}\ \emph {et~al.}(2019)\citenamefont {Heyl}, \citenamefont {Hauke},\ and\ \citenamefont {Zoller}}]{Heyl2019}%
  \BibitemOpen
  \bibfield  {author} {\bibinfo {author} {\bibfnamefont {M.}~\bibnamefont {Heyl}}, \bibinfo {author} {\bibfnamefont {P.}~\bibnamefont {Hauke}},\ and\ \bibinfo {author} {\bibfnamefont {P.}~\bibnamefont {Zoller}},\ }\bibfield  {title} {\bibinfo {title} {Quantum localization bounds trotter errors in digital quantum simulation},\ }\bibfield  {journal} {\bibinfo  {journal} {Science Advances}\ }\textbf {\bibinfo {volume} {5}},\ \href {https://doi.org/10.1126/sciadv.aau8342} {10.1126/sciadv.aau8342} (\bibinfo {year} {2019})\BibitemShut {NoStop}%
\bibitem [{\citenamefont {Wurtz}\ and\ \citenamefont {Love}(2022)}]{Wurtz2022}%
  \BibitemOpen
  \bibfield  {author} {\bibinfo {author} {\bibfnamefont {J.}~\bibnamefont {Wurtz}}\ and\ \bibinfo {author} {\bibfnamefont {P.~J.}\ \bibnamefont {Love}},\ }\bibfield  {title} {\bibinfo {title} {Counterdiabaticity and the quantum approximate optimization algorithm},\ }\href {https://doi.org/10.22331/q-2022-01-27-635} {\bibfield  {journal} {\bibinfo  {journal} {Quantum}\ }\textbf {\bibinfo {volume} {6}},\ \bibinfo {pages} {635} (\bibinfo {year} {2022})}\BibitemShut {NoStop}%
\bibitem [{\citenamefont {Kivlichan}\ \emph {et~al.}(2020)\citenamefont {Kivlichan}, \citenamefont {Gidney}, \citenamefont {Berry}, \citenamefont {Wiebe}, \citenamefont {McClean}, \citenamefont {Sun}, \citenamefont {Jiang}, \citenamefont {Rubin}, \citenamefont {Fowler}, \citenamefont {Aspuru-Guzik}, \citenamefont {Neven},\ and\ \citenamefont {Babbush}}]{Kivlichan2020}%
  \BibitemOpen
  \bibfield  {author} {\bibinfo {author} {\bibfnamefont {I.~D.}\ \bibnamefont {Kivlichan}}, \bibinfo {author} {\bibfnamefont {C.}~\bibnamefont {Gidney}}, \bibinfo {author} {\bibfnamefont {D.~W.}\ \bibnamefont {Berry}}, \bibinfo {author} {\bibfnamefont {N.}~\bibnamefont {Wiebe}}, \bibinfo {author} {\bibfnamefont {J.}~\bibnamefont {McClean}}, \bibinfo {author} {\bibfnamefont {W.}~\bibnamefont {Sun}}, \bibinfo {author} {\bibfnamefont {Z.}~\bibnamefont {Jiang}}, \bibinfo {author} {\bibfnamefont {N.}~\bibnamefont {Rubin}}, \bibinfo {author} {\bibfnamefont {A.}~\bibnamefont {Fowler}}, \bibinfo {author} {\bibfnamefont {A.}~\bibnamefont {Aspuru-Guzik}}, \bibinfo {author} {\bibfnamefont {H.}~\bibnamefont {Neven}},\ and\ \bibinfo {author} {\bibfnamefont {R.}~\bibnamefont {Babbush}},\ }\bibfield  {title} {\bibinfo {title} {Improved fault-tolerant quantum simulation of condensed-phase correlated electrons via trotterization},\ }\href@noop {} {\bibfield  {journal} {\bibinfo  {journal} {Quantum}\ }\textbf {\bibinfo
  {volume} {4}},\ \bibinfo {pages} {296} (\bibinfo {year} {2020})}\BibitemShut {NoStop}%
\bibitem [{\citenamefont {Zhao}\ \emph {et~al.}(2023)\citenamefont {Zhao}, \citenamefont {Bukov}, \citenamefont {Heyl},\ and\ \citenamefont {Moessner}}]{Hongzheng2023}%
  \BibitemOpen
  \bibfield  {author} {\bibinfo {author} {\bibfnamefont {H.}~\bibnamefont {Zhao}}, \bibinfo {author} {\bibfnamefont {M.}~\bibnamefont {Bukov}}, \bibinfo {author} {\bibfnamefont {M.}~\bibnamefont {Heyl}},\ and\ \bibinfo {author} {\bibfnamefont {R.}~\bibnamefont {Moessner}},\ }\bibfield  {title} {\bibinfo {title} {Making trotterization adaptive and energy-self-correcting for nisq devices and beyond},\ }\href {https://doi.org/10.1103/PRXQuantum.4.030319} {\bibfield  {journal} {\bibinfo  {journal} {PRX Quantum}\ }\textbf {\bibinfo {volume} {4}},\ \bibinfo {pages} {030319} (\bibinfo {year} {2023})}\BibitemShut {NoStop}%
\bibitem [{\citenamefont {Akila}\ \emph {et~al.}(2016)\citenamefont {Akila}, \citenamefont {Waltner}, \citenamefont {Gutkin},\ and\ \citenamefont {Guhr}}]{Akila2016}%
  \BibitemOpen
  \bibfield  {author} {\bibinfo {author} {\bibfnamefont {M.}~\bibnamefont {Akila}}, \bibinfo {author} {\bibfnamefont {D.}~\bibnamefont {Waltner}}, \bibinfo {author} {\bibfnamefont {B.}~\bibnamefont {Gutkin}},\ and\ \bibinfo {author} {\bibfnamefont {T.}~\bibnamefont {Guhr}},\ }\bibfield  {title} {\bibinfo {title} {Particle-time duality in the kicked ising spin chain},\ }\href {https://doi.org/10.1088/1751-8113/49/37/375101} {\bibfield  {journal} {\bibinfo  {journal} {Journal of Physics A: Mathematical and Theoretical}\ }\textbf {\bibinfo {volume} {49}},\ \bibinfo {pages} {375101} (\bibinfo {year} {2016})}\BibitemShut {NoStop}%
\bibitem [{\citenamefont {Bertini}\ \emph {et~al.}(2018)\citenamefont {Bertini}, \citenamefont {Kos},\ and\ \citenamefont {Prosen}}]{Bertini2018}%
  \BibitemOpen
  \bibfield  {author} {\bibinfo {author} {\bibfnamefont {B.}~\bibnamefont {Bertini}}, \bibinfo {author} {\bibfnamefont {P.}~\bibnamefont {Kos}},\ and\ \bibinfo {author} {\bibfnamefont {T.~c.~v.}\ \bibnamefont {Prosen}},\ }\bibfield  {title} {\bibinfo {title} {Exact spectral form factor in a minimal model of many-body quantum chaos},\ }\href {https://doi.org/10.1103/PhysRevLett.121.264101} {\bibfield  {journal} {\bibinfo  {journal} {Phys. Rev. Lett.}\ }\textbf {\bibinfo {volume} {121}},\ \bibinfo {pages} {264101} (\bibinfo {year} {2018})}\BibitemShut {NoStop}%
\bibitem [{\citenamefont {Bertini}\ \emph {et~al.}(2019)\citenamefont {Bertini}, \citenamefont {Kos},\ and\ \citenamefont {Prosen}}]{Bertini2019}%
  \BibitemOpen
  \bibfield  {author} {\bibinfo {author} {\bibfnamefont {B.}~\bibnamefont {Bertini}}, \bibinfo {author} {\bibfnamefont {P.}~\bibnamefont {Kos}},\ and\ \bibinfo {author} {\bibfnamefont {T.~c.~v.}\ \bibnamefont {Prosen}},\ }\bibfield  {title} {\bibinfo {title} {Entanglement spreading in a minimal model of maximal many-body quantum chaos},\ }\href {https://doi.org/10.1103/PhysRevX.9.021033} {\bibfield  {journal} {\bibinfo  {journal} {Phys. Rev. X}\ }\textbf {\bibinfo {volume} {9}},\ \bibinfo {pages} {021033} (\bibinfo {year} {2019})}\BibitemShut {NoStop}%
\bibitem [{\citenamefont {Gopalakrishnan}\ and\ \citenamefont {Lamacraft}(2019)}]{Gopalakrishnan2019}%
  \BibitemOpen
  \bibfield  {author} {\bibinfo {author} {\bibfnamefont {S.}~\bibnamefont {Gopalakrishnan}}\ and\ \bibinfo {author} {\bibfnamefont {A.}~\bibnamefont {Lamacraft}},\ }\bibfield  {title} {\bibinfo {title} {Unitary circuits of finite depth and infinite width from quantum channels},\ }\bibfield  {journal} {\bibinfo  {journal} {Physical Review B}\ }\textbf {\bibinfo {volume} {100}},\ \href {https://doi.org/10.1103/physrevb.100.064309} {10.1103/physrevb.100.064309} (\bibinfo {year} {2019})\BibitemShut {NoStop}%
\bibitem [{\citenamefont {Altman}\ \emph {et~al.}(2021)\citenamefont {Altman}, \citenamefont {Brown}, \citenamefont {Carleo}, \citenamefont {Carr}, \citenamefont {Demler}, \citenamefont {Chin}, \citenamefont {DeMarco}, \citenamefont {Economou}, \citenamefont {Eriksson}, \citenamefont {Fu}, \citenamefont {Greiner}, \citenamefont {Hazzard}, \citenamefont {Hulet}, \citenamefont {Koll\'ar}, \citenamefont {Lev}, \citenamefont {Lukin}, \citenamefont {Ma}, \citenamefont {Mi}, \citenamefont {Misra}, \citenamefont {Monroe}, \citenamefont {Murch}, \citenamefont {Nazario}, \citenamefont {Ni}, \citenamefont {Potter}, \citenamefont {Roushan}, \citenamefont {Saffman}, \citenamefont {Schleier-Smith}, \citenamefont {Siddiqi}, \citenamefont {Simmonds}, \citenamefont {Singh}, \citenamefont {Spielman}, \citenamefont {Temme}, \citenamefont {Weiss}, \citenamefont {Vu\ifmmode \check{c}\else \v{c}\fi{}kovi\ifmmode~\acute{c}\else \'{c}\fi{}}, \citenamefont {Vuleti\ifmmode~\acute{c}\else \'{c}\fi{}}, \citenamefont {Ye},\ and\
  \citenamefont {Zwierlein}}]{AltmanRev2021}%
  \BibitemOpen
  \bibfield  {author} {\bibinfo {author} {\bibfnamefont {E.}~\bibnamefont {Altman}}, \bibinfo {author} {\bibfnamefont {K.~R.}\ \bibnamefont {Brown}}, \bibinfo {author} {\bibfnamefont {G.}~\bibnamefont {Carleo}}, \bibinfo {author} {\bibfnamefont {L.~D.}\ \bibnamefont {Carr}}, \bibinfo {author} {\bibfnamefont {E.}~\bibnamefont {Demler}}, \bibinfo {author} {\bibfnamefont {C.}~\bibnamefont {Chin}}, \bibinfo {author} {\bibfnamefont {B.}~\bibnamefont {DeMarco}}, \bibinfo {author} {\bibfnamefont {S.~E.}\ \bibnamefont {Economou}}, \bibinfo {author} {\bibfnamefont {M.~A.}\ \bibnamefont {Eriksson}}, \bibinfo {author} {\bibfnamefont {K.-M.~C.}\ \bibnamefont {Fu}}, \bibinfo {author} {\bibfnamefont {M.}~\bibnamefont {Greiner}}, \bibinfo {author} {\bibfnamefont {K.~R.}\ \bibnamefont {Hazzard}}, \bibinfo {author} {\bibfnamefont {R.~G.}\ \bibnamefont {Hulet}}, \bibinfo {author} {\bibfnamefont {A.~J.}\ \bibnamefont {Koll\'ar}}, \bibinfo {author} {\bibfnamefont {B.~L.}\ \bibnamefont {Lev}}, \bibinfo {author} {\bibfnamefont
  {M.~D.}\ \bibnamefont {Lukin}}, \bibinfo {author} {\bibfnamefont {R.}~\bibnamefont {Ma}}, \bibinfo {author} {\bibfnamefont {X.}~\bibnamefont {Mi}}, \bibinfo {author} {\bibfnamefont {S.}~\bibnamefont {Misra}}, \bibinfo {author} {\bibfnamefont {C.}~\bibnamefont {Monroe}}, \bibinfo {author} {\bibfnamefont {K.}~\bibnamefont {Murch}}, \bibinfo {author} {\bibfnamefont {Z.}~\bibnamefont {Nazario}}, \bibinfo {author} {\bibfnamefont {K.-K.}\ \bibnamefont {Ni}}, \bibinfo {author} {\bibfnamefont {A.~C.}\ \bibnamefont {Potter}}, \bibinfo {author} {\bibfnamefont {P.}~\bibnamefont {Roushan}}, \bibinfo {author} {\bibfnamefont {M.}~\bibnamefont {Saffman}}, \bibinfo {author} {\bibfnamefont {M.}~\bibnamefont {Schleier-Smith}}, \bibinfo {author} {\bibfnamefont {I.}~\bibnamefont {Siddiqi}}, \bibinfo {author} {\bibfnamefont {R.}~\bibnamefont {Simmonds}}, \bibinfo {author} {\bibfnamefont {M.}~\bibnamefont {Singh}}, \bibinfo {author} {\bibfnamefont {I.}~\bibnamefont {Spielman}}, \bibinfo {author} {\bibfnamefont {K.}~\bibnamefont
  {Temme}}, \bibinfo {author} {\bibfnamefont {D.~S.}\ \bibnamefont {Weiss}}, \bibinfo {author} {\bibfnamefont {J.}~\bibnamefont {Vu\ifmmode \check{c}\else \v{c}\fi{}kovi\ifmmode~\acute{c}\else \'{c}\fi{}}}, \bibinfo {author} {\bibfnamefont {V.}~\bibnamefont {Vuleti\ifmmode~\acute{c}\else \'{c}\fi{}}}, \bibinfo {author} {\bibfnamefont {J.}~\bibnamefont {Ye}},\ and\ \bibinfo {author} {\bibfnamefont {M.}~\bibnamefont {Zwierlein}},\ }\bibfield  {title} {\bibinfo {title} {Quantum simulators: Architectures and opportunities},\ }\href {https://doi.org/10.1103/PRXQuantum.2.017003} {\bibfield  {journal} {\bibinfo  {journal} {PRX Quantum}\ }\textbf {\bibinfo {volume} {2}},\ \bibinfo {pages} {017003} (\bibinfo {year} {2021})}\BibitemShut {NoStop}%
\bibitem [{\citenamefont {Ho}\ and\ \citenamefont {Choi}(2022)}]{Ho2022}%
  \BibitemOpen
  \bibfield  {author} {\bibinfo {author} {\bibfnamefont {W.~W.}\ \bibnamefont {Ho}}\ and\ \bibinfo {author} {\bibfnamefont {S.}~\bibnamefont {Choi}},\ }\bibfield  {title} {\bibinfo {title} {Exact emergent quantum state designs from quantum chaotic dynamics},\ }\href {http://dx.doi.org/10.1103/PhysRevLett.128.060601} {\bibfield  {journal} {\bibinfo  {journal} {Physical Review Letters}\ }\textbf {\bibinfo {volume} {128}} (\bibinfo {year} {2022})}\BibitemShut {NoStop}%
\bibitem [{\citenamefont {Haake}\ \emph {et~al.}(1987)\citenamefont {Haake}, \citenamefont {Kuś},\ and\ \citenamefont {Scharf}}]{Haake1987}%
  \BibitemOpen
  \bibfield  {author} {\bibinfo {author} {\bibfnamefont {F.}~\bibnamefont {Haake}}, \bibinfo {author} {\bibfnamefont {M.}~\bibnamefont {Kuś}},\ and\ \bibinfo {author} {\bibfnamefont {R.}~\bibnamefont {Scharf}},\ }\bibfield  {title} {\bibinfo {title} {Classical and quantum chaos for a kicked top},\ }\href {https://doi.org/10.1007/bf01303727} {\bibfield  {journal} {\bibinfo  {journal} {Zeitschrift f\"{u}r Physik B Condensed Matter}\ }\textbf {\bibinfo {volume} {65}},\ \bibinfo {pages} {381–395} (\bibinfo {year} {1987})}\BibitemShut {NoStop}%
\bibitem [{\citenamefont {Chaudhury}\ \emph {et~al.}(2009)\citenamefont {Chaudhury}, \citenamefont {Smith}, \citenamefont {Anderson}, \citenamefont {Ghose},\ and\ \citenamefont {Jessen}}]{Chaudhury2009}%
  \BibitemOpen
  \bibfield  {author} {\bibinfo {author} {\bibfnamefont {S.}~\bibnamefont {Chaudhury}}, \bibinfo {author} {\bibfnamefont {A.}~\bibnamefont {Smith}}, \bibinfo {author} {\bibfnamefont {B.~E.}\ \bibnamefont {Anderson}}, \bibinfo {author} {\bibfnamefont {S.}~\bibnamefont {Ghose}},\ and\ \bibinfo {author} {\bibfnamefont {P.~S.}\ \bibnamefont {Jessen}},\ }\bibfield  {title} {\bibinfo {title} {Quantum signatures of chaos in a kicked top},\ }\href {https://doi.org/10.1038/nature08396} {\bibfield  {journal} {\bibinfo  {journal} {Nature}\ }\textbf {\bibinfo {volume} {461}},\ \bibinfo {pages} {768–771} (\bibinfo {year} {2009})}\BibitemShut {NoStop}%
\bibitem [{\citenamefont {Neill}\ \emph {et~al.}(2016)\citenamefont {Neill}, \citenamefont {Roushan}, \citenamefont {Fang}, \citenamefont {Chen}, \citenamefont {Kolodrubetz}, \citenamefont {Chen}, \citenamefont {Megrant}, \citenamefont {Barends}, \citenamefont {Campbell}, \citenamefont {Chiaro}, \citenamefont {Dunsworth}, \citenamefont {Jeffrey}, \citenamefont {Kelly}, \citenamefont {Mutus}, \citenamefont {O’Malley}, \citenamefont {Quintana}, \citenamefont {Sank}, \citenamefont {Vainsencher}, \citenamefont {Wenner}, \citenamefont {White}, \citenamefont {Polkovnikov},\ and\ \citenamefont {Martinis}}]{Neill2016}%
  \BibitemOpen
  \bibfield  {author} {\bibinfo {author} {\bibfnamefont {C.}~\bibnamefont {Neill}}, \bibinfo {author} {\bibfnamefont {P.}~\bibnamefont {Roushan}}, \bibinfo {author} {\bibfnamefont {M.}~\bibnamefont {Fang}}, \bibinfo {author} {\bibfnamefont {Y.}~\bibnamefont {Chen}}, \bibinfo {author} {\bibfnamefont {M.}~\bibnamefont {Kolodrubetz}}, \bibinfo {author} {\bibfnamefont {Z.}~\bibnamefont {Chen}}, \bibinfo {author} {\bibfnamefont {A.}~\bibnamefont {Megrant}}, \bibinfo {author} {\bibfnamefont {R.}~\bibnamefont {Barends}}, \bibinfo {author} {\bibfnamefont {B.}~\bibnamefont {Campbell}}, \bibinfo {author} {\bibfnamefont {B.}~\bibnamefont {Chiaro}}, \bibinfo {author} {\bibfnamefont {A.}~\bibnamefont {Dunsworth}}, \bibinfo {author} {\bibfnamefont {E.}~\bibnamefont {Jeffrey}}, \bibinfo {author} {\bibfnamefont {J.}~\bibnamefont {Kelly}}, \bibinfo {author} {\bibfnamefont {J.}~\bibnamefont {Mutus}}, \bibinfo {author} {\bibfnamefont {P.~J.~J.}\ \bibnamefont {O’Malley}}, \bibinfo {author} {\bibfnamefont {C.}~\bibnamefont
  {Quintana}}, \bibinfo {author} {\bibfnamefont {D.}~\bibnamefont {Sank}}, \bibinfo {author} {\bibfnamefont {A.}~\bibnamefont {Vainsencher}}, \bibinfo {author} {\bibfnamefont {J.}~\bibnamefont {Wenner}}, \bibinfo {author} {\bibfnamefont {T.~C.}\ \bibnamefont {White}}, \bibinfo {author} {\bibfnamefont {A.}~\bibnamefont {Polkovnikov}},\ and\ \bibinfo {author} {\bibfnamefont {J.~M.}\ \bibnamefont {Martinis}},\ }\bibfield  {title} {\bibinfo {title} {Ergodic dynamics and thermalization in an isolated quantum system},\ }\href {https://doi.org/10.1038/nphys3830} {\bibfield  {journal} {\bibinfo  {journal} {Nature Physics}\ }\textbf {\bibinfo {volume} {12}},\ \bibinfo {pages} {1037–1041} (\bibinfo {year} {2016})}\BibitemShut {NoStop}%
\bibitem [{\citenamefont {Sieberer}\ \emph {et~al.}(2019)\citenamefont {Sieberer}, \citenamefont {Olsacher}, \citenamefont {Elben}, \citenamefont {Heyl}, \citenamefont {Hauke}, \citenamefont {Haake},\ and\ \citenamefont {Zoller}}]{Sieberer2019}%
  \BibitemOpen
  \bibfield  {author} {\bibinfo {author} {\bibfnamefont {L.~M.}\ \bibnamefont {Sieberer}}, \bibinfo {author} {\bibfnamefont {T.}~\bibnamefont {Olsacher}}, \bibinfo {author} {\bibfnamefont {A.}~\bibnamefont {Elben}}, \bibinfo {author} {\bibfnamefont {M.}~\bibnamefont {Heyl}}, \bibinfo {author} {\bibfnamefont {P.}~\bibnamefont {Hauke}}, \bibinfo {author} {\bibfnamefont {F.}~\bibnamefont {Haake}},\ and\ \bibinfo {author} {\bibfnamefont {P.}~\bibnamefont {Zoller}},\ }\bibfield  {title} {\bibinfo {title} {Digital quantum simulation, trotter errors, and quantum chaos of the kicked top},\ }\bibfield  {journal} {\bibinfo  {journal} {npj Quantum Information}\ }\textbf {\bibinfo {volume} {5}},\ \href {https://doi.org/10.1038/s41534-019-0192-5} {10.1038/s41534-019-0192-5} (\bibinfo {year} {2019})\BibitemShut {NoStop}%
\bibitem [{\citenamefont {Hahn}(1950)}]{Hahn1950}%
  \BibitemOpen
  \bibfield  {author} {\bibinfo {author} {\bibfnamefont {E.~L.}\ \bibnamefont {Hahn}},\ }\bibfield  {title} {\bibinfo {title} {Spin echoes},\ }\href {https://doi.org/10.1103/PhysRev.80.580} {\bibfield  {journal} {\bibinfo  {journal} {Phys. Rev.}\ }\textbf {\bibinfo {volume} {80}},\ \bibinfo {pages} {580} (\bibinfo {year} {1950})}\BibitemShut {NoStop}%
\bibitem [{\citenamefont {Tanner}\ and\ \citenamefont {Stejskal}(1968)}]{Tanner1968}%
  \BibitemOpen
  \bibfield  {author} {\bibinfo {author} {\bibfnamefont {J.~E.}\ \bibnamefont {Tanner}}\ and\ \bibinfo {author} {\bibfnamefont {E.~O.}\ \bibnamefont {Stejskal}},\ }\bibfield  {title} {\bibinfo {title} {Restricted self-diffusion of protons in colloidal systems by the pulsed-gradient, spin-echo method},\ }\href {https://doi.org/10.1063/1.1670306} {\bibfield  {journal} {\bibinfo  {journal} {The Journal of Chemical Physics}\ }\textbf {\bibinfo {volume} {49}},\ \bibinfo {pages} {1768–1777} (\bibinfo {year} {1968})}\BibitemShut {NoStop}%
\bibitem [{\citenamefont {Weichselbaumer}\ \emph {et~al.}(2020)\citenamefont {Weichselbaumer}, \citenamefont {Zens}, \citenamefont {Zollitsch}, \citenamefont {Brandt}, \citenamefont {Rotter}, \citenamefont {Gross},\ and\ \citenamefont {Huebl}}]{Hans2020}%
  \BibitemOpen
  \bibfield  {author} {\bibinfo {author} {\bibfnamefont {S.}~\bibnamefont {Weichselbaumer}}, \bibinfo {author} {\bibfnamefont {M.}~\bibnamefont {Zens}}, \bibinfo {author} {\bibfnamefont {C.~W.}\ \bibnamefont {Zollitsch}}, \bibinfo {author} {\bibfnamefont {M.~S.}\ \bibnamefont {Brandt}}, \bibinfo {author} {\bibfnamefont {S.}~\bibnamefont {Rotter}}, \bibinfo {author} {\bibfnamefont {R.}~\bibnamefont {Gross}},\ and\ \bibinfo {author} {\bibfnamefont {H.}~\bibnamefont {Huebl}},\ }\bibfield  {title} {\bibinfo {title} {Echo trains in pulsed electron spin resonance of a strongly coupled spin ensemble},\ }\href {https://link.aps.org/doi/10.1103/PhysRevLett.125.137701} {\bibfield  {journal} {\bibinfo  {journal} {Phys. Rev. Lett.}\ }\textbf {\bibinfo {volume} {125}},\ \bibinfo {pages} {137701} (\bibinfo {year} {2020})}\BibitemShut {NoStop}%
\bibitem [{\citenamefont {Lasek}\ \emph {et~al.}(2023)\citenamefont {Lasek}, \citenamefont {Lepage}, \citenamefont {Zhang}, \citenamefont {Ferrus},\ and\ \citenamefont {Barnes}}]{Lasek2023}%
  \BibitemOpen
  \bibfield  {author} {\bibinfo {author} {\bibfnamefont {A.}~\bibnamefont {Lasek}}, \bibinfo {author} {\bibfnamefont {H.~V.}\ \bibnamefont {Lepage}}, \bibinfo {author} {\bibfnamefont {K.}~\bibnamefont {Zhang}}, \bibinfo {author} {\bibfnamefont {T.}~\bibnamefont {Ferrus}},\ and\ \bibinfo {author} {\bibfnamefont {C.~H.~W.}\ \bibnamefont {Barnes}},\ }\bibfield  {title} {\bibinfo {title} {Pulse-controlled qubit in semiconductor double quantum dots},\ }\bibfield  {journal} {\bibinfo  {journal} {Scientific Reports}\ }\textbf {\bibinfo {volume} {13}},\ \href {https://doi.org/10.1038/s41598-023-47405-0} {10.1038/s41598-023-47405-0} (\bibinfo {year} {2023})\BibitemShut {NoStop}%
\bibitem [{\citenamefont {Sambe}(1973)}]{Sambe1973}%
  \BibitemOpen
  \bibfield  {author} {\bibinfo {author} {\bibfnamefont {H.}~\bibnamefont {Sambe}},\ }\bibfield  {title} {\bibinfo {title} {Steady states and quasienergies of a quantum-mechanical system in an oscillating field},\ }\href {https://doi.org/10.1103/PhysRevA.7.2203} {\bibfield  {journal} {\bibinfo  {journal} {Phys. Rev. A}\ }\textbf {\bibinfo {volume} {7}},\ \bibinfo {pages} {2203} (\bibinfo {year} {1973})}\BibitemShut {NoStop}%
\bibitem [{\citenamefont {Eckardt}\ and\ \citenamefont {Anisimovas}(2015)}]{Eckardt2015}%
  \BibitemOpen
  \bibfield  {author} {\bibinfo {author} {\bibfnamefont {A.}~\bibnamefont {Eckardt}}\ and\ \bibinfo {author} {\bibfnamefont {E.}~\bibnamefont {Anisimovas}},\ }\bibfield  {title} {\bibinfo {title} {High-frequency approximation for periodically driven quantum systems from a floquet-space perspective},\ }\href {https://doi.org/10.1088/1367-2630/17/9/093039} {\bibfield  {journal} {\bibinfo  {journal} {New Journal of Physics}\ }\textbf {\bibinfo {volume} {17}},\ \bibinfo {pages} {093039} (\bibinfo {year} {2015})}\BibitemShut {NoStop}%
\bibitem [{\citenamefont {Hewitt}\ and\ \citenamefont {Hewitt}(1979)}]{Hewitt1979}%
  \BibitemOpen
  \bibfield  {author} {\bibinfo {author} {\bibfnamefont {E.}~\bibnamefont {Hewitt}}\ and\ \bibinfo {author} {\bibfnamefont {R.~E.}\ \bibnamefont {Hewitt}},\ }\bibfield  {title} {\bibinfo {title} {The {Gibbs-Wilbraham} phenomenon: An episode in fourier analysis},\ }\href@noop {} {\bibfield  {journal} {\bibinfo  {journal} {Arch. Hist. Exact Sci.}\ }\textbf {\bibinfo {volume} {21}},\ \bibinfo {pages} {129} (\bibinfo {year} {1979})}\BibitemShut {NoStop}%
\bibitem [{\citenamefont {Walsh}(1923)}]{Walsh1923}%
  \BibitemOpen
  \bibfield  {author} {\bibinfo {author} {\bibfnamefont {J.~L.}\ \bibnamefont {Walsh}},\ }\bibfield  {title} {\bibinfo {title} {A closed set of normal orthogonal functions},\ }\href {https://doi.org/10.2307/2387224} {\bibfield  {journal} {\bibinfo  {journal} {American Journal of Mathematics}\ }\textbf {\bibinfo {volume} {45}},\ \bibinfo {pages} {5} (\bibinfo {year} {1923})}\BibitemShut {NoStop}%
\bibitem [{\citenamefont {Fine}(1949)}]{Fine1949}%
  \BibitemOpen
  \bibfield  {author} {\bibinfo {author} {\bibfnamefont {N.~J.}\ \bibnamefont {Fine}},\ }\bibfield  {title} {\bibinfo {title} {On the walsh functions},\ }\href {https://doi.org/10.1090/s0002-9947-1949-0032833-2} {\bibfield  {journal} {\bibinfo  {journal} {Transactions of the American Mathematical Society}\ }\textbf {\bibinfo {volume} {65}},\ \bibinfo {pages} {372–414} (\bibinfo {year} {1949})}\BibitemShut {NoStop}%
\bibitem [{\citenamefont {Golubov}\ \emph {et~al.}(2012)\citenamefont {Golubov}, \citenamefont {Efimov},\ and\ \citenamefont {Skvortsov}}]{golubov2012}%
  \BibitemOpen
  \bibfield  {author} {\bibinfo {author} {\bibfnamefont {B.}~\bibnamefont {Golubov}}, \bibinfo {author} {\bibfnamefont {A.}~\bibnamefont {Efimov}},\ and\ \bibinfo {author} {\bibfnamefont {V.}~\bibnamefont {Skvortsov}},\ }\href@noop {} {\emph {\bibinfo {title} {Walsh series and transforms: theory and applications}}},\ Vol.~\bibinfo {volume} {64}\ (\bibinfo  {publisher} {Springer Science \& Business Media},\ \bibinfo {year} {2012})\BibitemShut {NoStop}%
\bibitem [{\citenamefont {Hayes}\ \emph {et~al.}(2011)\citenamefont {Hayes}, \citenamefont {Khodjasteh}, \citenamefont {Viola},\ and\ \citenamefont {Biercuk}}]{Hayes2011}%
  \BibitemOpen
  \bibfield  {author} {\bibinfo {author} {\bibfnamefont {D.}~\bibnamefont {Hayes}}, \bibinfo {author} {\bibfnamefont {K.}~\bibnamefont {Khodjasteh}}, \bibinfo {author} {\bibfnamefont {L.}~\bibnamefont {Viola}},\ and\ \bibinfo {author} {\bibfnamefont {M.~J.}\ \bibnamefont {Biercuk}},\ }\bibfield  {title} {\bibinfo {title} {Reducing sequencing complexity in dynamical quantum error suppression by walsh modulation},\ }\bibfield  {journal} {\bibinfo  {journal} {Physical Review A}\ }\textbf {\bibinfo {volume} {84}},\ \href {https://doi.org/10.1103/physreva.84.062323} {10.1103/physreva.84.062323} (\bibinfo {year} {2011})\BibitemShut {NoStop}%
\bibitem [{\citenamefont {Shukla}\ and\ \citenamefont {Vedula}(2023)}]{Shukla2023}%
  \BibitemOpen
  \bibfield  {author} {\bibinfo {author} {\bibfnamefont {A.}~\bibnamefont {Shukla}}\ and\ \bibinfo {author} {\bibfnamefont {P.}~\bibnamefont {Vedula}},\ }\bibfield  {title} {\bibinfo {title} {A hybrid classical-quantum algorithm for solution of nonlinear ordinary differential equations},\ }\href {https://doi.org/10.1016/j.amc.2022.127708} {\bibfield  {journal} {\bibinfo  {journal} {Applied Mathematics and Computation}\ }\textbf {\bibinfo {volume} {442}},\ \bibinfo {pages} {127708} (\bibinfo {year} {2023})}\BibitemShut {NoStop}%
\bibitem [{\citenamefont {Zylberman}\ and\ \citenamefont {Debbasch}(2024)}]{zylberman2024}%
  \BibitemOpen
  \bibfield  {author} {\bibinfo {author} {\bibfnamefont {J.}~\bibnamefont {Zylberman}}\ and\ \bibinfo {author} {\bibfnamefont {F.}~\bibnamefont {Debbasch}},\ }\bibfield  {title} {\bibinfo {title} {Efficient quantum state preparation with walsh series},\ }\href {https://doi.org/10.1103/PhysRevA.109.042401} {\bibfield  {journal} {\bibinfo  {journal} {Phys. Rev. A}\ }\textbf {\bibinfo {volume} {109}},\ \bibinfo {pages} {042401} (\bibinfo {year} {2024})}\BibitemShut {NoStop}%
\bibitem [{\citenamefont {Pagano}\ \emph {et~al.}(2024)\citenamefont {Pagano}, \citenamefont {M\"{u}ller}, \citenamefont {Calarco}, \citenamefont {Montangero},\ and\ \citenamefont {Rembold}}]{Pagano2024}%
  \BibitemOpen
  \bibfield  {author} {\bibinfo {author} {\bibfnamefont {A.}~\bibnamefont {Pagano}}, \bibinfo {author} {\bibfnamefont {M.~M.}\ \bibnamefont {M\"{u}ller}}, \bibinfo {author} {\bibfnamefont {T.}~\bibnamefont {Calarco}}, \bibinfo {author} {\bibfnamefont {S.}~\bibnamefont {Montangero}},\ and\ \bibinfo {author} {\bibfnamefont {P.}~\bibnamefont {Rembold}},\ }\href {https://doi.org/10.48550/ARXIV.2405.20889} {\bibinfo {title} {The role of bases in quantum optimal control}} (\bibinfo {year} {2024})\BibitemShut {NoStop}%
\bibitem [{\citenamefont {Votto}\ \emph {et~al.}(2024)\citenamefont {Votto}, \citenamefont {Zeiher},\ and\ \citenamefont {Vermersch}}]{Votto2024}%
  \BibitemOpen
  \bibfield  {author} {\bibinfo {author} {\bibfnamefont {M.}~\bibnamefont {Votto}}, \bibinfo {author} {\bibfnamefont {J.}~\bibnamefont {Zeiher}},\ and\ \bibinfo {author} {\bibfnamefont {B.}~\bibnamefont {Vermersch}},\ }\bibfield  {title} {\bibinfo {title} {Universal quantum processors in spin systems via robust local pulse sequences},\ }\href {https://doi.org/10.22331/q-2024-10-29-1513} {\bibfield  {journal} {\bibinfo  {journal} {Quantum}\ }\textbf {\bibinfo {volume} {8}},\ \bibinfo {pages} {1513} (\bibinfo {year} {2024})}\BibitemShut {NoStop}%
\bibitem [{\citenamefont {Zhihua}\ and\ \citenamefont {Qishan}(1983)}]{Zhihua1983}%
  \BibitemOpen
  \bibfield  {author} {\bibinfo {author} {\bibfnamefont {L.}~\bibnamefont {Zhihua}}\ and\ \bibinfo {author} {\bibfnamefont {Z.}~\bibnamefont {Qishan}},\ }\bibfield  {title} {\bibinfo {title} {Ordering of walsh functions},\ }\href {https://doi.org/10.1109/TEMC.1983.304153} {\bibfield  {journal} {\bibinfo  {journal} {IEEE Transactions on Electromagnetic Compatibility}\ }\textbf {\bibinfo {volume} {EMC-25}},\ \bibinfo {pages} {115} (\bibinfo {year} {1983})}\BibitemShut {NoStop}%
\bibitem [{\citenamefont {Nyquist}(1928)}]{Nyquist1928}%
  \BibitemOpen
  \bibfield  {author} {\bibinfo {author} {\bibfnamefont {H.}~\bibnamefont {Nyquist}},\ }\bibfield  {title} {\bibinfo {title} {Certain topics in telegraph transmission theory},\ }\href {https://doi.org/10.1109/t-aiee.1928.5055024} {\bibfield  {journal} {\bibinfo  {journal} {Transactions of the American Institute of Electrical Engineers}\ }\textbf {\bibinfo {volume} {47}},\ \bibinfo {pages} {617–644} (\bibinfo {year} {1928})}\BibitemShut {NoStop}%
\bibitem [{Note1()}]{Note1}%
  \BibitemOpen
  \bibinfo {note} {$h_x$ is a dimensionless quantity in natural units since it is the amplitude of a delta function drive. $h_z$ is dimensionful since it just multiplies some time independent term.}\BibitemShut {Stop}%
\bibitem [{\citenamefont {Goldman}\ and\ \citenamefont {Dalibard}(2014)}]{Goldman2014}%
  \BibitemOpen
  \bibfield  {author} {\bibinfo {author} {\bibfnamefont {N.}~\bibnamefont {Goldman}}\ and\ \bibinfo {author} {\bibfnamefont {J.}~\bibnamefont {Dalibard}},\ }\bibfield  {title} {\bibinfo {title} {Periodically driven quantum systems: Effective hamiltonians and engineered gauge fields},\ }\href {https://doi.org/10.1103/PhysRevX.4.031027} {\bibfield  {journal} {\bibinfo  {journal} {Phys. Rev. X}\ }\textbf {\bibinfo {volume} {4}},\ \bibinfo {pages} {031027} (\bibinfo {year} {2014})}\BibitemShut {NoStop}%
\bibitem [{\citenamefont {Wang}\ \emph {et~al.}(2024)\citenamefont {Wang}, \citenamefont {Zhu}, \citenamefont {Li}, \citenamefont {Li}, \citenamefont {Viola}, \citenamefont {Cooper},\ and\ \citenamefont {Cappellaro}}]{Wang2024}%
  \BibitemOpen
  \bibfield  {author} {\bibinfo {author} {\bibfnamefont {G.}~\bibnamefont {Wang}}, \bibinfo {author} {\bibfnamefont {Y.}~\bibnamefont {Zhu}}, \bibinfo {author} {\bibfnamefont {B.}~\bibnamefont {Li}}, \bibinfo {author} {\bibfnamefont {C.}~\bibnamefont {Li}}, \bibinfo {author} {\bibfnamefont {L.}~\bibnamefont {Viola}}, \bibinfo {author} {\bibfnamefont {A.}~\bibnamefont {Cooper}},\ and\ \bibinfo {author} {\bibfnamefont {P.}~\bibnamefont {Cappellaro}},\ }\bibfield  {title} {\bibinfo {title} {Digital noise spectroscopy with a quantum sensor},\ }\href {https://doi.org/10.1088/2058-9565/ad3846} {\bibfield  {journal} {\bibinfo  {journal} {Quantum Science and Technology}\ }\textbf {\bibinfo {volume} {9}},\ \bibinfo {pages} {035006} (\bibinfo {year} {2024})}\BibitemShut {NoStop}%
\bibitem [{\citenamefont {Schindler}\ and\ \citenamefont {Bukov}(2024{\natexlab{a}})}]{Schindler2024}%
  \BibitemOpen
  \bibfield  {author} {\bibinfo {author} {\bibfnamefont {P.~M.}\ \bibnamefont {Schindler}}\ and\ \bibinfo {author} {\bibfnamefont {M.}~\bibnamefont {Bukov}},\ }\bibfield  {title} {\bibinfo {title} {Counterdiabatic driving for periodically driven systems},\ }\href {http://dx.doi.org/10.1103/PhysRevLett.133.123402} {\bibfield  {journal} {\bibinfo  {journal} {Physical Review Letters}\ }\textbf {\bibinfo {volume} {133}} (\bibinfo {year} {2024}{\natexlab{a}})}\BibitemShut {NoStop}%
\bibitem [{\citenamefont {Schindler}\ and\ \citenamefont {Bukov}(2024{\natexlab{b}})}]{schindler2024geometric}%
  \BibitemOpen
  \bibfield  {author} {\bibinfo {author} {\bibfnamefont {P.~M.}\ \bibnamefont {Schindler}}\ and\ \bibinfo {author} {\bibfnamefont {M.}~\bibnamefont {Bukov}},\ }\bibfield  {title} {\bibinfo {title} {Geometric floquet theory},\ }\href {https://arxiv.org/abs/2410.07029} {\bibfield  {journal} {\bibinfo  {journal} {arXiv preprint arXiv:2410.07029}\ } (\bibinfo {year} {2024}{\natexlab{b}})}\BibitemShut {NoStop}%
\bibitem [{\citenamefont {Ferreiro-Vélez}\ \emph {et~al.}(2024)\citenamefont {Ferreiro-Vélez}, \citenamefont {Iriarte-Zendoia}, \citenamefont {Ban},\ and\ \citenamefont {Chen}}]{xichen2024}%
  \BibitemOpen
  \bibfield  {author} {\bibinfo {author} {\bibfnamefont {J.}~\bibnamefont {Ferreiro-Vélez}}, \bibinfo {author} {\bibfnamefont {I.}~\bibnamefont {Iriarte-Zendoia}}, \bibinfo {author} {\bibfnamefont {Y.}~\bibnamefont {Ban}},\ and\ \bibinfo {author} {\bibfnamefont {X.}~\bibnamefont {Chen}},\ }\href {https://doi.org/10.48550/ARXIV.2407.20957} {\bibinfo {title} {Shortcuts for adiabatic and variational algorithms in molecular simulation}} (\bibinfo {year} {2024})\BibitemShut {NoStop}%
\bibitem [{\citenamefont {Sun}\ \emph {et~al.}(2022)\citenamefont {Sun}, \citenamefont {Chandarana}, \citenamefont {Xin},\ and\ \citenamefont {Chen}}]{Sun2022}%
  \BibitemOpen
  \bibfield  {author} {\bibinfo {author} {\bibfnamefont {D.}~\bibnamefont {Sun}}, \bibinfo {author} {\bibfnamefont {P.}~\bibnamefont {Chandarana}}, \bibinfo {author} {\bibfnamefont {Z.-H.}\ \bibnamefont {Xin}},\ and\ \bibinfo {author} {\bibfnamefont {X.}~\bibnamefont {Chen}},\ }\bibfield  {title} {\bibinfo {title} {Optimizing counterdiabaticity by variational quantum circuits},\ }\bibfield  {journal} {\bibinfo  {journal} {Philosophical Transactions of the Royal Society A: Mathematical, Physical and Engineering Sciences}\ }\textbf {\bibinfo {volume} {380}},\ \href {https://doi.org/10.1098/rsta.2021.0282} {10.1098/rsta.2021.0282} (\bibinfo {year} {2022})\BibitemShut {NoStop}%
\bibitem [{\citenamefont {Georges}\ \emph {et~al.}(2025)\citenamefont {Georges}, \citenamefont {Berntson}, \citenamefont {S\"{u}nderhauf},\ and\ \citenamefont {Ivanov}}]{Georges2025}%
  \BibitemOpen
  \bibfield  {author} {\bibinfo {author} {\bibfnamefont {T.~N.}\ \bibnamefont {Georges}}, \bibinfo {author} {\bibfnamefont {B.~K.}\ \bibnamefont {Berntson}}, \bibinfo {author} {\bibfnamefont {C.}~\bibnamefont {S\"{u}nderhauf}},\ and\ \bibinfo {author} {\bibfnamefont {A.~V.}\ \bibnamefont {Ivanov}},\ }\bibfield  {title} {\bibinfo {title} {Pauli decomposition via the fast walsh-hadamard transform},\ }\href {https://doi.org/10.1088/1367-2630/adb44d} {\bibfield  {journal} {\bibinfo  {journal} {New Journal of Physics}\ }\textbf {\bibinfo {volume} {27}},\ \bibinfo {pages} {033004} (\bibinfo {year} {2025})}\BibitemShut {NoStop}%
\bibitem [{\citenamefont {Cornwell}(1997)}]{cornwell1997}%
  \BibitemOpen
  \bibfield  {author} {\bibinfo {author} {\bibfnamefont {J.~F.}\ \bibnamefont {Cornwell}},\ }\href@noop {} {\emph {\bibinfo {title} {Group Theory in Physics, Volume 1}}}\ (\bibinfo  {publisher} {Academic Press},\ \bibinfo {address} {New York},\ \bibinfo {year} {1997})\BibitemShut {NoStop}%
\bibitem [{\citenamefont {Shannon}(1949)}]{Shannon1949}%
  \BibitemOpen
  \bibfield  {author} {\bibinfo {author} {\bibfnamefont {C.}~\bibnamefont {Shannon}},\ }\bibfield  {title} {\bibinfo {title} {Communication in the presence of noise},\ }\href {https://doi.org/10.1109/JRPROC.1949.232969} {\bibfield  {journal} {\bibinfo  {journal} {Proceedings of the IRE}\ }\textbf {\bibinfo {volume} {37}},\ \bibinfo {pages} {10} (\bibinfo {year} {1949})}\BibitemShut {NoStop}%
\bibitem [{Note2()}]{Note2}%
  \BibitemOpen
  \bibinfo {note} {The anti-symmetry follows from the fact that $f_{ab}+f_{ba}= \DOTSB \sum@ \slimits@ _{\alpha > \beta } W_{\alpha a}W_{\beta b}+ \DOTSB \sum@ \slimits@ _{\beta > \alpha }W_{\alpha a}W_{\beta b} = \DOTSB \sum@ \slimits@ _{\alpha ,\beta } W_{\alpha a} W_{\beta b} \propto \delta _{a0}\delta _{b0}$ which is zero for all non-zero commutator terms.}\BibitemShut {Stop}%
\bibitem [{\citenamefont {Blanes}\ \emph {et~al.}(2009)\citenamefont {Blanes}, \citenamefont {Casas}, \citenamefont {Oteo},\ and\ \citenamefont {Ros}}]{BLANES2009}%
  \BibitemOpen
  \bibfield  {author} {\bibinfo {author} {\bibfnamefont {S.}~\bibnamefont {Blanes}}, \bibinfo {author} {\bibfnamefont {F.}~\bibnamefont {Casas}}, \bibinfo {author} {\bibfnamefont {J.}~\bibnamefont {Oteo}},\ and\ \bibinfo {author} {\bibfnamefont {J.}~\bibnamefont {Ros}},\ }\bibfield  {title} {\bibinfo {title} {The magnus expansion and some of its applications},\ }\href {https://doi.org/https://doi.org/10.1016/j.physrep.2008.11.001} {\bibfield  {journal} {\bibinfo  {journal} {Physics Reports}\ }\textbf {\bibinfo {volume} {470}},\ \bibinfo {pages} {151} (\bibinfo {year} {2009})}\BibitemShut {NoStop}%
\bibitem [{\citenamefont {Bravyi}\ \emph {et~al.}(2011)\citenamefont {Bravyi}, \citenamefont {DiVincenzo},\ and\ \citenamefont {Loss}}]{Bravyi2011}%
  \BibitemOpen
  \bibfield  {author} {\bibinfo {author} {\bibfnamefont {S.}~\bibnamefont {Bravyi}}, \bibinfo {author} {\bibfnamefont {D.~P.}\ \bibnamefont {DiVincenzo}},\ and\ \bibinfo {author} {\bibfnamefont {D.}~\bibnamefont {Loss}},\ }\bibfield  {title} {\bibinfo {title} {Schrieffer–wolff transformation for quantum many-body systems},\ }\href {https://doi.org/10.1016/j.aop.2011.06.004} {\bibfield  {journal} {\bibinfo  {journal} {Annals of Physics}\ }\textbf {\bibinfo {volume} {326}},\ \bibinfo {pages} {2793–2826} (\bibinfo {year} {2011})}\BibitemShut {NoStop}%
\bibitem [{\citenamefont {Sandoval‐Santana}\ \emph {et~al.}(2019)\citenamefont {Sandoval‐Santana}, \citenamefont {Ibarra‐Sierra}, \citenamefont {Cardoso}, \citenamefont {Kunold}, \citenamefont {Roman‐Taboada},\ and\ \citenamefont {Naumis}}]{SandovalSantana2019}%
  \BibitemOpen
  \bibfield  {author} {\bibinfo {author} {\bibfnamefont {J.~C.}\ \bibnamefont {Sandoval‐Santana}}, \bibinfo {author} {\bibfnamefont {V.~G.}\ \bibnamefont {Ibarra‐Sierra}}, \bibinfo {author} {\bibfnamefont {J.~L.}\ \bibnamefont {Cardoso}}, \bibinfo {author} {\bibfnamefont {A.}~\bibnamefont {Kunold}}, \bibinfo {author} {\bibfnamefont {P.}~\bibnamefont {Roman‐Taboada}},\ and\ \bibinfo {author} {\bibfnamefont {G.}~\bibnamefont {Naumis}},\ }\bibfield  {title} {\bibinfo {title} {Method for finding the exact effective hamiltonian of time‐driven quantum systems},\ }\bibfield  {journal} {\bibinfo  {journal} {Annalen der Physik}\ }\textbf {\bibinfo {volume} {531}},\ \href {https://doi.org/10.1002/andp.201900035} {10.1002/andp.201900035} (\bibinfo {year} {2019})\BibitemShut {NoStop}%
\end{thebibliography}%

\clearpage

\begin{widetext}
\begin{center}
\textbf{\large Supplementary Material for:\\[0.3em]
Walsh–Floquet Theory of Periodic Kick Drives}
\end{center}
\end{widetext}

\setcounter{section}{0}
\setcounter{figure}{0}
\renewcommand{\thesection}{S\arabic{section}}
\renewcommand{\thesubsection}{S\arabic{section}.\arabic{subsection}}
\renewcommand{\theequation}{S\arabic{equation}}
\renewcommand{\thefigure}{S\arabic{figure}}
\renewcommand{\thetable}{S\arabic{table}}

\let\addcontentsline\origsaddcontentsline

\begingroup
\makeatletter
\renewcommand{\addcontentsline}[3]{%
  \addtocontents{#1}{\protect\contentsline{#2}{#3}{\thepage}{}}%
}
\makeatother
\tableofcontents
\endgroup
\vspace{0.5 cm}

This supplementary material is organized as follows. In the first two sections, we review the formalism behind this paper; Sec.~\ref{sec:Floquet} discusses the basics of the Floquet theorem, and Sec.~\ref{sec:Walsh} shows the construction of the Walsh basis alongside its relationship with the discrete Fourier transform. In Secs.~\ref{sec:Time} and \ref{sec:Truncation}, we discuss the formalism for Floquet extended space and the definition of the time translation generator. Secs.~\ref{sec:Square}, \ref{sec:Scaling} and \ref{sec:Expansion} add to the results in the main text by exploring the case of a square drive, studying error scaling and detailing the convention for the up-down kick drive. In Secs.~\ref{sec:Strong} and \ref{sec:Inverse-Frequency}, we introduce further interesting features related to the Walsh basis in Floquet; polaritonic states composed of Walsh functions alongside an inverse-frequency expansion in the Walsh basis. 

\section{Floquet Theorem}\label{sec:Floquet}
Floquet's theorem states that there exists a frame in which the dynamics of a periodically driven system are time independent \cite{Bukov2015}. The Floquet Hamiltonian, 
\begin{equation}
    \hat{H}_F[t_0] = \hat{P}^\dagger(t) \hat{H}(t) \hat{P}(t) - i \hat{P}^\dagger(t) \partial_{t} \hat{P}(t),
\end{equation}
determines the full dynamics starting from a time $t_0$ in the rotating frame defined via the micromotion operator, \begin{equation}
\hat{P}(t) \equiv\hat{P}(t,t_0) = \exp\left(-i\hat{K}[t_0](t)\right).
\end{equation} The eigenvalues (quasienergies), $\varepsilon_n$, and eigenstates (Floquet modes), $\ket{u_n(t_0)}$, of $\hat{H}_F[t_0]$ can be directly used to find the time evolution of the solutions to the Schrödinger equation as $\ket{\psi_n(t)} = e^{-i\varepsilon_n t} \ket{u_n(t)}$ which is an alternative statement of Floquet's theorem.  We need to characterise the time-dependence of the modes to understand the response of the system to the periodic drive. The Floquet modes' evolution is governed by the micromotion operator: $\ket{u_n(t)} = \hat{P}(t,t_0)\ket{u_n(t_0)}$. Since the time evolution of $\ket{u_n(t)}$ is not given by the Hamiltonian, the Floquet modes are sometimes equivalently written as $\ket{u_n[t]}$ which emphasises the Floquet gauge dependence. 

In the high frequency regime, the van-Vleck expansion for the kick operator, $\hat{K}_{\text{vV}}(t)$, in terms of the Fourier modes of the Hamiltonian, $\hat{H}_l$, is given by \cite{Goldman2014, Eckardt2015, Bukov2015}
\begin{equation}
    \begin{aligned}
    \hat{K}_{\text{vV}}(t) &= \sum_{l \neq 0} \frac{\hat{H}_l}{il \omega}e^{il\omega t} + \mathcal{O}\left(\omega^{-2}\right)\\ &= \int_{0}^{t} \text{d} s \, \hat{H}(s) + \mathcal{O}\left(\omega^{-2}\right).
    \end{aligned}
    \label{eqn:vVleck}
\end{equation}

This gives an approximation of the kick operator to understand the dynamics governing the evolution of the Floquet modes. This expansion demonstrates that at high-frequency, the response is approximately the integral of the drive.

\section{Walsh Basis Details}\label{sec:Walsh}
Formally, the discrete version of the Walsh basis is constructed from the columns (or equivalently rows) of the character table, $\chi$, of the group $(\mathbb{Z}_2)^n$ for $n\in \mathbb{N}$ \cite{Walsh1923, golubov2012}. This can be expressed in terms of the Hadamard matrices $\mathbb{H}_n$ of order $2^n$ (Fig.~2). The resultant character table for the direct product of two groups $F = G \times H$ is the tensor product of the two character tables \cite{cornwell1997}: $\chi_F= \chi_G \otimes \chi_H$.
Since $\chi_{\mathbb{Z}_2}=\mathbb{H}_2$, the Walsh basis elements are constructed from different orders of Hadamard matrices that come from taking the tensor product of $\mathbb{H}_2$ with itself: 
\begin{equation}
    F= (\mathbb{Z}_2)^n \qquad \Rightarrow \qquad \chi_F = \bigotimes^{n}_{k=1} \mathbb{H}_2 = \mathbb{H}_{2^n}.
\end{equation}
The orthogonality of the Walsh basis is guaranteed by the Great Orthogonality Theorem \cite{cornwell1997}. The natural ordering of Walsh functions used throughout this paper is that which comes from the ordering of the columns of the Hadamard matrix. There are also several other orderings, such as sequency ordering, which organizes them by the number of roots \cite{Zhihua1983}. 

The difference in representation of functions in the Walsh vs.~Fourier basis are best understood through an intermediary: the discrete Fourier basis has the same basis functions as the Fourier, but is only defined on discrete times like the Walsh. When $N$, the number of basis elements, is a power of $2$, the discrete Fourier basis can be unitarily mapped to the Walsh basis exactly at any finite level of truncation. In this sense, they are equivalent.

\begin{figure*}[t]
    \centering
    \includegraphics[scale=1]{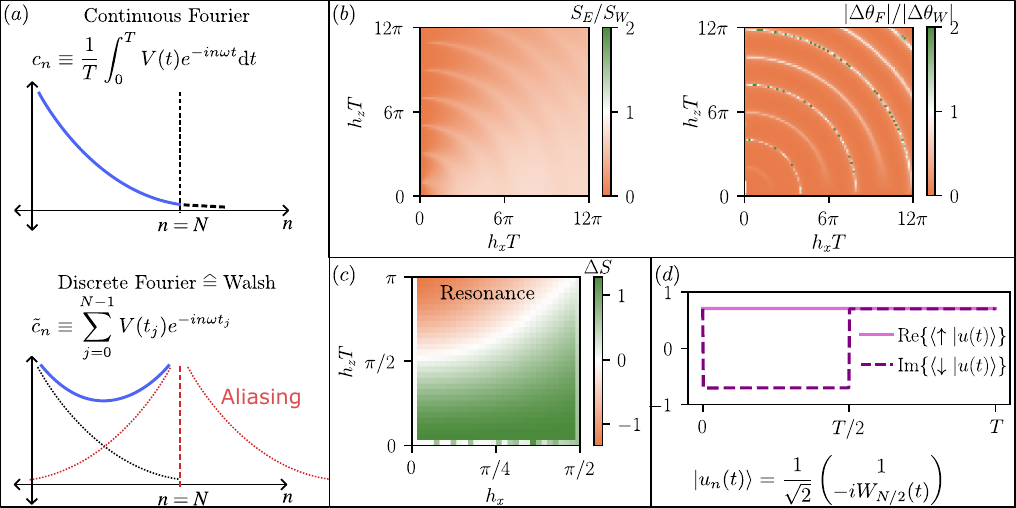}
    \caption{
    (a) plots of the coefficients for the continuous Fourier transform, $c_n$, and discrete Fourier transform, $\tilde{c}_n$, in the case of a square wave. The difference arises due to the nature of the discrete time, which leads to the aliasing effect in the discrete Fourier coefficients. 
    (b) equivalent comparison between localization and error as in Fig.~5(b) from the main text but for a square wave drive. 
    (c) participation entropy for the spin-down component of a two-level system state for a kick drive which suffers from delocalization due to a Fourier polariton (spin-photon entanglement) near $h_z=\omega/2$. 
    (d) example of a Walsh polariton where the two spin components differ by a factor of a Walsh function. Not shown real and imaginary parts vanish.
    }
    \label{fig:Endmatter}
\end{figure*}

For driving period $T$, the discrete Fourier coefficients, $\tilde{c}_n$ (Eq.~\eqref{eqn:ctwiddle}), are given by the discrete convolution of the Fourier coefficients, $c_n$, and a delta comb with spacing $\Delta t = T/N$ as shown in Fig.~\ref{fig:Endmatter}(a). Based on the definitions of the inner products and the discrete Fourier transform,

\begin{align}
    \tilde{c}_m &= \frac{1}{N} \sum_{n=0}^{N-1} f(t_n) e^{-im\omega t_n} \label{eqn:ctwiddle}\\
    &=  \frac{1}{N} \int_0^T \text{d} t\, \text{M}(t) f(t) e^{-im\omega t}\\
    &= \frac{T}{N}\sum_{m'} c_{m-m'} \tilde{\text{M}}_{m'}\equiv \tilde{\text{M}}_m * c_m,
\end{align}
where $\text{M}(t) = \sum_{n=0}^{N-1} \delta(t-t_n)$ is the delta comb and $t_n=nT/N$. This is the convolution theorem for discrete Fourier coefficients. From a physical perspective, the distortion of the spectrum can also be understood as aliasing as described by the Nyquist-Shannon theorem \cite{Shannon1949}; our sampling frequency, $f_s = N/T$, is lower than twice the maximum frequency in the spectrum giving distortion. We illustrate the relationship between $c_m$ and $\tilde{c}_m$ in Fig.~\ref{fig:Endmatter}(a) in the case of a square wave drive ($c_m \sim 1/m$).

In the continuous Fourier case, we implement a hard cutoff at $n=N$. However, in the case of the discrete Fourier (and Walsh basis), the entire spectrum is mapped to within the cutoff. This is shown in Fig.~\ref{fig:Endmatter} by the red aliasing copy which adds together with infinitely many copies to give the (blue) discrete spectrum in the first zone. 

\section{Time Translation Generator}\label{sec:Time}

\subsection{Definition and Derivation}
The operator of interest is the derivative operator of the Schrödinger equation. Crucially, this operator is also the generator of time translations. When we work in a finite, non-analytic basis, such as the Walsh basis, the derivative and the generator of translations do not have the same spectrum for a finite truncation, so they must be distinguished from one another. This is seen partly due to the failure of the typical Taylor series argument, that establishes the correspondence between the two operators, which cannot be employed for the elements of the non-analytic Walsh basis.

To yield the correct results when we compare with other bases, we must explicitly use the generator of time translations. We construct the operator for discrete time translations of size $\Delta t$ and then scale appropriately to get an approximation to a continuum operator for arbitrary translations.

To understand time translations, we first need to define the vector space the operators act on. The Walsh basis and the discrete Fourier basis are defined using only a discrete set of times $t_j = j T/N$. Essentially, we take some function defined on the continuum, $f(t)$, and then map it onto a vector where the components represent the different times:
\begin{equation}
f(t) \mapsto
    \mathbf{f} = \begin{pmatrix}
f(t_0) \\
f(t_1) \\
f(t_2) \\
\vdots \\
f(t_{N-1})
\end{pmatrix}
\end{equation}

\begin{figure}[t!]
    \centering
    \includegraphics[width=\linewidth]{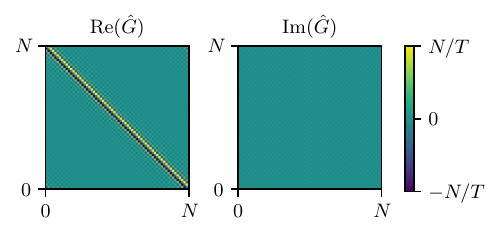}
    \caption{A plot of the real and imaginary matrix elements of $\hat{G}$, cf.~Eq.~\eqref{eq:T_and_G}, when written in real space (i.e., not in Fourier space) for $N=64$. While the real part appears similar to the derivative, as expected, there is a broadening of non-zero values around the central diagonal due to the finite discretization.
    }
    \label{fig:realspacederiv}
\end{figure}

To generate time translations, we use a \textit{discrete time translation operator}. This is the cyclic left-shift matrix given by 
\begin{equation}
    \mathbf{T} = \begin{pmatrix}
0 & 1 & 0 & \cdots & 0 & 0 \\
0 & 0 & 1 & \cdots & 0 & 0 \\
0 & 0 & 0 & \cdots & 0 & 0 \\
\vdots & \vdots & \vdots & \ddots & \vdots & \vdots \\
0 & 0 & 0 & \cdots & 0 & 1 \\
1 & 0 & 0 & \cdots & 0 & 0
\end{pmatrix},
\label{eqn:translation}
\end{equation}
which performs the transformation $f(t_n) \mapsto f_T(t_n)=f(t_{n+1})=f(t_n+T/N)$. We employ periodic boundary conditions due to the periodic basis. Taking the matrix logarithm, we find the generator of time translations:
\begin{equation}
\label{eq:T_and_G}
    \mathbf{T} = \exp( \hat{G} T/N) 
    \quad\longrightarrow\quad \hat{G} = \frac{N}{T}\log(\mathbf{T}). 
\end{equation}
This operator $\hat{G}$ is a good approximation to the true generator of continuous time translations in the infinite Hilbert space since their truncated spectra are identical. Visualizing $\hat{G}$ as a matrix, it has the form shown in Fig.~\ref{fig:realspacederiv}. 

A guess for an antisymmetric derivative matrix could use finite differences like $\sim f(t_n) - f(t_{n-2})$, which $\hat{G}$ does have in addition to further contributions (See Fig.~\ref{fig:realspacederiv}). However, unlike the naive guess, the spectrum of $\hat{G}$ exactly matches that of the true derivative in the continuum at any finite truncation.

\begin{figure}
    \centering
    \includegraphics[scale=1]{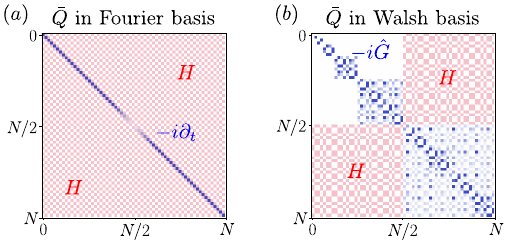}
    \caption{Schematic of the quasienergy operator in the Fourier and Walsh bases for a periodic up-down kick drive. The matrix elements of the drive are shown in red, whereas the terms from the time-translation operator are shown in blue. (a) the quasienergy operator in the Fourier basis has a uniform all-to-all coupling via the kick drive elements. (b) in the Walsh basis, the time-translation operator is block diagonal.}
    \label{fig:appendixQmatrix}
\end{figure}

To convert the time translation operator into the space of the Walsh basis or into the discrete Fourier basis, we use their respective transformation matrices (Hadamard and cyclic exponentials). Combining the generator of time translations with the Hamiltonian in the Walsh basis to form $\bar{Q}=H-i\hat{G}$, we arrive at the matrix shown schematically in Fig.~\ref{fig:appendixQmatrix}. 

In Fig.~\ref{fig:appendixQmatrix}(b), we show the discrete time translation generator in blue in the Walsh basis. In contrast to the Fourier basis, it is block diagonal in natural ordering, but with relatively sparse blocks. The symmetry operation corresponding to the blocks acts on subsets of Walsh functions with the same smallest unit cell such that translations permute between these subsets.

The spectrum is identical to the truncated continuum operator which is just integer multiples of $\omega$. We find that all the odd multiples (i.e $\pm\omega, \pm 3\omega, \pm5\omega, \dots$) are associated with the large block in the lower right corner for the Walsh basis. For the diagonal blocks higher up, the spectrum of the block is odd integers multiplied by $2^b$ where $b$ labels the block number starting from zero in the bottom right.

\subsection{Representation in Walsh Basis}
As discussed in the main text, the rows of the Hadamard matrices are the Walsh basis functions. The translation operator by $T/4$ in the Walsh basis for $N=4$ is given by 

\begin{equation}
    \mathbf{T}_\text{Walsh}=\exp(\hat{G}T/4)=\begin{pmatrix}
        1& 0 & 0 & 0\\
        0 & -1 & 0 & 0 \\
        0 & 0 & 0 & -1 \\
        0 & 0& 1 & 0 \\
    \end{pmatrix}.
    \label{eqn:translationWalsh}
\end{equation}

This form makes sense since both $W_0$ and $W_1$ from Fig.~2(a) map to themselves (up to a minus sign) under translation by $T/4$, and $W_2$ and $W_3$ map to each other up to a minus sign. This is numerically calculated for larger $N$ by converting Eq.~\eqref{eqn:translation} into the Walsh basis using the Hadamard matrices: $ \frac{1}{N} \mathbb{H}_{N}\mathbf{T} \mathbb{H}_{N}=\mathbf{T}_\text{Walsh}$. 

Rewriting $T=2\pi/\omega$, and taking the matrix logarithm of Eq.~\eqref{eqn:translationWalsh}, we find
\begin{equation}
    \hat{G} = \frac{2\omega}{\pi}\begin{pmatrix}
        0& 0 & 0 & 0\\
        0 & i\pi & 0 & 0 \\
        0 & 0 & 0 & -\frac{\pi}{2} \\
        0 & 0& \frac{\pi}{2} & 0 \\
    \end{pmatrix}= \begin{pmatrix}
        0& 0 & 0 & 0\\
        0 & i2\omega & 0 & 0 \\
        0 & 0 & 0 & -\omega \\
        0 & 0& \omega & 0 \\
    \end{pmatrix}.
    \label{eqn:derivWalshN=4}
\end{equation}

As discussed earlier, this has the correct spectrum of multiples of $\omega$ for $i\partial_t \equiv i\hat{G}$ as we expect in the full continuum for Fourier. The block diagonal structure is also already evident, and it is also preserved for larger $N$.

\subsection{Symmetrization of Walsh Eigenvalues}

Due to the definition of the translation operator in the Walsh basis, the quasienergy spectrum is inherently asymmetric for the Walsh basis. We can see the asymmetry of the spectrum directly in the case of the resonances in Fig.~4(a) and explicitly in the construction from Eq.~\eqref{eqn:derivWalshN=4}.

The case $N=4$, (for $-i\partial_t\equiv -i\hat{G}$), yields eigenvalues $(-2\omega, -\omega, 0, \omega)$. In principle, we can symmetrize this spectrum by eliminating the lowest eigenvalue which can be performed for any $N$. For the drive elements, this means removing some rows and columns associated with $-2\omega$, thus removing $W_3$ from the basis. Under symmetrization, the translation generator of Eq.~\eqref{eqn:derivWalshN=4} becomes
\begin{equation}
\begin{pmatrix}
        0& 0 & 0 & 0\\
        0 & i2\omega & 0 & 0 \\
        0 & 0 & 0 & -\omega \\
        0 & 0& \omega & 0 \\
    \end{pmatrix} \rightarrow \begin{pmatrix}
        0 & 0 & 0\\
        0 & 0 & -\omega \\
         0& \omega & 0 \\
    \end{pmatrix},
\end{equation}
which now has a symmetric spectrum $(-i\omega, 0, i\omega).$

Alternatively, we have the freedom to also take $( -\omega, 0, \omega, 2\omega)$ since either $\pm 2\omega$ produce a minus sign from discrete translation. One could devise increasingly contrived symmetrization methods such as averaging out the effects of these two different cases to see if that produces better results. In fine-tuned problems, these techniques can be used to obtain more accurate solutions; however, they come at the cost of removing the completeness alongside the group structure and symmetry inherent in the Walsh basis. Hence, in this work, we have generally opted not to apply this procedure. 

\section{Truncation of Quasienergy Operator}\label{sec:Truncation}
The quasienergy phases, $\theta=\varepsilon T$, can be calculated from the eigenvalues of the quasienergy operator $\bar{Q}$. As discussed in the main text, this operator is infinite in extent, since we can couple states with arbitrarily many photons with an arbitrarily high-order process. The eigenvalues of the matrix can be numerically approximated by performing exact diagonalisation on a truncated version of the matrix.

In the case of the Fourier basis, we pick a finite number of basis elements for the space $\mathcal{L}_\odot$ by truncating to $\mathcal{L}^{(2M+1)}_\odot = \{e^{-iM\omega t}, \dots, e^{-i\omega t}, 1, e^{i\omega t}, \dots, e^{iM\omega t} \}$. This is described as ``truncating the frequency lattice" since we only take a finite number of ``sites" corresponding to different multiples of $\omega$. We then calculate the matrix elements of $\bar{Q}$ corresponding to this finite set of functions (using Eq.~2) and store them in an array (Fig.~\ref{fig:appendixQmatrix}) which we exactly diagonalize. The procedure is identical for the Walsh basis except we expand in the basis $\mathcal{L}^{(N)}_\odot = \{W_0(t), W_1(t), \dots, W_{N-1}(t)\}$.

\section{Square Drive Localization vs. Error}\label{sec:Square}
The basis of square functions, the Walsh basis, does not outpeform the Fourier for a square wave drive in the shape of $W_{N/2}(t)$. In Fig.~\ref{fig:Endmatter}(b), we compare the error to the localization as in Fig.~5(c) in the main text. For a square wave drive, the quasienergy error is at least an order of magnitude smaller for Fourier than Walsh. This surprising result follows from the response being approximately the integral of the drive at high frequency (Eq.~\eqref{eqn:vVleck}); a square wave drive has a triangle wave response, which requires fewer harmonics in the Fourier basis than the Walsh. 

The participation entropy correctly predicts the hierarchy of Walsh vs.~Fourier, and both error and entropy possess a similar ring structure (Fig.~\ref{fig:Endmatter}(b)); however, the maxima of the entropy rings are at odd multiples of $\pi$, whereas the maxima of the quasienergy rings are at even multiples of $\pi$. This discrepancy demonstrates the shortcomings of the localization measure; the Fourier spectrum for even multiples of $\pi$ consists of a single mode which results in a highly localized signature, suggesting the error is far better than it actually is. 

\section{Scaling of Errors with Number of Modes}\label{sec:Scaling}
We now discuss the scaling of the quasienergy errors with respect to various parameters in the two bases in the many-body case (Eq.~3) and single-particle case (Eq.~3 with $J=0$). We pick some illustrative values of $h_x, h_z$ such that we are firmly in the high-frequency regime. The two key parameters of interest for our systems are the driving frequency, $\omega$, and the number of elements we keep before truncation, $N$.

\begin{figure}
    \centering
    \includegraphics[width=\linewidth]{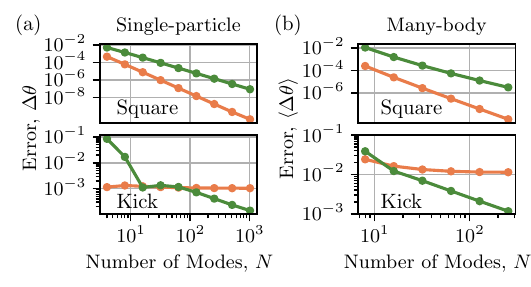}
    \caption{Scaling of the quasienergy phase error in the first Floquet Brillouin zone, $\Delta \theta$, with number of modes, $N$ for Walsh (green) and Fourier (orange). (a) shows the error for a single-particle, and (b) shows the error for the many-body Hamiltonian (Eq.~3) with $J=1$ and $L=3$. The top row shows results for a square wave drive where both show power law convergence with $N$. The parameters are $\omega=50$ with $h_z=1$ and $h_x=6$. The bottom row is the error for the up-down kick drive where the convergence of the solutions is not self-evident and Walsh outperforms Fourier by at least an order of magnitude. The parameters are $\omega=10$ with $h_z=5.5$ and $h_x=0.2$.
    }
    \label{fig:errorscaling}
\end{figure}

For the case of a square drive, we find that the quasienergy error in each case scales as 
\begin{equation}
    \Delta \theta_{\text{Fourier}} \sim N^{-3} \omega^{-3},\qquad
\Delta \theta_{\text{Walsh}} \sim N^{-2} \omega^{-3}.
\end{equation}
In general, the $\omega$ scaling of the error is universal for sufficiently weak drives since then the problem becomes equivalent to second-order perturbation theory about a diagonal derivative term with energy difference $\sim \omega$. For the square wave drive (Fig.~\ref{fig:errorscaling}(a)), we show that the Walsh fails to outperform the Fourier basis in this particular case, even with increasing $N$. This observation is in agreement with Fig.~\ref{fig:Endmatter}(b) and the associated single-particle localization theory that we explore in the main text. For the strongly driven regime of a kick drive, the scaling in $N$ is non-universal, and also does not appear to have a power law form in general, as shown in the lower row of Fig.~\ref{fig:errorscaling}, but we do see that the Walsh continues to outperform Fourier over a range of $N$. 

\begin{figure}
    \centering
    \includegraphics[width=\linewidth]{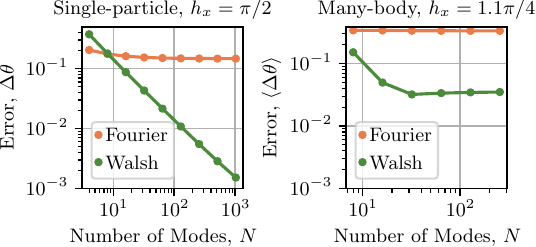}
    \caption{Convergence analysis for the same parameters in Fig.~4(a) for different kick strengths, $h_x$. While strong kicks ($h_x \sim \pi/2$) converge for the single-particle case (left panel), for many-body (right panel), even moderately strong kicks fail to converge. The large error is due to few states affected strongly by resonances as can be seen in Fig.~4(a).}
    \label{fig:nonconverge}
\end{figure}

In the case of particularly strong kicking and low frequency in the many-body case, it is possible to have many resonances; these spectral features are a challenge to capture, and they act as an obstacle to convergence in both the Walsh and Fourier bases for a few states. We observe this lack of convergence for $\theta$ with increasing $N$ for the parameters of Fig.~4(a) in the main text; the corresponding scaling is shown in the right panel of Fig.~\ref{fig:nonconverge}. However, parts of the spectrum mostly unaffected by resonances can show order of magnitude smaller errors, but will not necessarily converge too. In contrast, the single-particle case converges for much stronger strong kicks (left panel of Fig.~\ref{fig:nonconverge}).

\section{Expansion Coefficients of Walsh Basis for the Up-Down Kick Drive}\label{sec:Expansion}
\begin{figure}[t!]
    \centering
    \includegraphics[width=\linewidth]{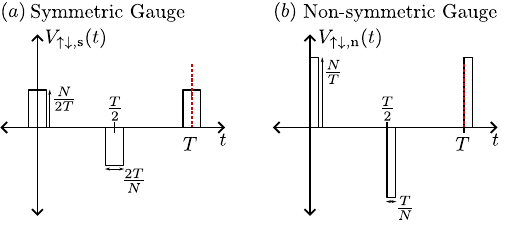}
    \caption{(a) symmetric convention and (b) non-symmetric convention for discretizing the up-down periodic kick drive ($V_{\uparrow \downarrow}$).}
    \label{fig:appendixkick}
\end{figure}

The up-down kick drive consists of a positive delta pulse at $t=0$ and a negative pulse at $t=T/2$ (see Fig.~1(a) in the main text). There are two conventions one can use to discretize this function illustrated in Fig.~\ref{fig:appendixkick}; we can either symmetrically split the peak at $t=0$ into two ``half" delta functions (symmetric, as in (a)), or shift the delta peak by $\delta t=\epsilon>0$ to the right (non-symmetric, as in (b)). From a Floquet perspective, these correspond to shifting the Floquet gauge~\cite{Bukov2015}, $t_0$, which affects none of our results concerning quasienergies. 

In the main text, we use the symmetrized convention because it has fewer non-zero coefficients; only Walsh functions with opposite signs at $t=0$ and $t=T/2$ have non-zero overlap with the symmetrized up-down kick. These are easiest to study using sequency ordering \cite{Zhihua1983}, since this makes direct reference to the number of roots which dictates the signs at the points of interest. Walsh functions with an odd sequency start and end with an opposite sign. Doubling the sequency places the functions reflected end to end. Hence, only Walsh functions which have double an odd sequency, $m_\text{allowed} = 4m+2$ where $m\in \mathbb{N}$, are allowed. Thus, we find that $V_{\uparrow \downarrow,\text{s}}(t)= 2 \sum_{m=0}^\infty \mathcal{W}_{4m+2}(t)$. 

For the non-symmetrized convention, there are twice as many modes in the expansion: $V_{\uparrow \downarrow, \text{n}}(t)= 2\left( \sum_{m=0}^\infty \mathcal{W}_{4m+1}(t)+\mathcal{W}_{4m+2}(t) \right).$ 

\section{Strong Spin-Photon Hybridization (Walsh Polariton)}\label{sec:Strong}
While in the main text we only show the spin-up component, the spin-down component can be quite different in character. This is shown in Fig.~\ref{fig:Endmatter}(d).

We describe the strong hybridisation of spin and photon degrees of freedom as a polariton in the system. On resonance, generically in the limit of arbitrarily weak driving, we observe polariton states. These lead to significant errors for the Walsh basis, since the photon is associated with a factor $e^{im\omega t}$ which is strongly delocalized for the Walsh. This means that while these resonances lead to relative translations between spin component Fourier modes, leaving the localization unchanged, they drastically change the behavior of the localization between the spin degrees of freedom in the Walsh basis.

In the limit of strong driving, it is also possible to get Walsh polaritons as we observed for the kick with strength $\pi/2$; the two different spin components have a relative Walsh function time dependence between them. This could be observed in experimental platforms such as digital quantum simulators or NISQ devices.

\section{Inverse-Frequency Expansion in the Walsh basis}\label{sec:Inverse-Frequency}

For a digital drive that takes constant values on certain intervals, we label the value of the Hamiltonian on time interval starting at $t_j$ by $H(t_{j})=H_{j}$. Thus, the unitary evolution operator in the case where we have $N=2^n$ intervals is given by
\begin{equation}
    U(T) = \prod^{\leftarrow}_{j} \exp\left( -\frac{iH_{j}T}{N}\right)\equiv \exp(-iH_{\text{eff}}T),
\end{equation}
where the product runs from earlier to later times, as indicated by the arrow. By using a generalized version of the Baker-Campbell-Hausdorff expansion (BCH) where we allow for arbitrary products, we can write an expansion for $H_\text{eff}$ in terms of $T$. For $N=2$, we recover the usual BCH expansion:
\begin{equation}
    H_{\text{eff}, N=2} = \frac{H_{0}+H_{1}}{2}- \frac{iT}{8}[H_{1},H_{0}]+\cdots\, .
\end{equation}

For the case of general $N$, we compound the formula by splitting the time intervals for $H_0$ and $H_1$ as well. In the real space (time basis), the effective Hamiltonian can be expressed as 
\begin{equation}
    H_{\text{eff},N}=\bar{H}-\frac{iT}{2N^2} \sum_{i>j} [H_{i},H_{j}]+ \mathcal{O} \left( \frac{T^2}{N^3} \right).
\end{equation}
This is the most natural way to frame the problem in the Walsh basis since it is inherently defined only at discrete times. 

To convert this into the Walsh basis, we perform a Walsh-Hadamard transformation of the Hamiltonian. We define the new coefficients as 
\begin{equation}
    H_{j}=H(t_j)=\sum_{m=0}^{N-1}h_{m}  W_{mj},
\end{equation}
where the symmetric Hadamard matrix is defined by $W_{mj} \equiv W_m(t_j)$. If we substitute this form in, we get the expansion
\begin{equation}
    H_{\text{eff},N}= h_0 + iT\sum_{a>b} f_{ab}[h_a,h_b] +\cdots,
\end{equation}
where
\begin{equation}
    f_{ab} = -\frac{1}{N^2}\sum_{\alpha > \beta} W_{\alpha a} W_{\beta b}
    \label{eqn:deffab}
\end{equation}
defines the coefficients. We restrict the sum using a redundancy to $f_{ab} = -f_{ba}$~\footnote{The anti-symmetry follows from the fact that $f_{ab}+f_{ba}= \sum_{\alpha> \beta} W_{\alpha a}W_{\beta b}+ \sum_{\beta > \alpha}W_{\alpha a}W_{\beta b} = \sum_{\alpha,\beta} W_{\alpha a} W_{\beta b} \propto \delta_{a0}\delta_{b0}$ which is zero for all non-zero commutator terms.} which cancels the factor of $1/2$  in the series expansion. The factor of $1/N^2$ is included in Eq.~\eqref{eqn:deffab} to make $f_{ab}$ $N$-independent; this allows us to tabulate values for $f_{ab}$ that can be used for any number of basis functions. The matrix $f_{ab}$ has a number of non-zero elements that scales linearly with the basis size, much like in the Floquet-Magnus expansion \cite{BLANES2009}. 

We tabulate some non-zero values of the matrix according to their sequency labels (in contrast to the natural ordering used in the main text) since then these coefficients hold true for all $N$.

\begin{table}[ht]
    \centering
    \begin{tabular}{>{\centering\arraybackslash}p{2.5cm} >{\centering\arraybackslash}p{6.1cm}}
    \textbf{$f_{ab}$} & \textbf{$(a,b)$ (Sequency Labels)} \\
    \hline
    $2^{-2}$ & (1, 0) \\
    $2^{-3}$ & (3,0), (2, 1) \\
    $2^{-4}$ & (7, 0), (6, 1), (5, 2), (4,3) \\
    $2^{-5}$ & (15, 0), (14,1), (13, 2), (12, 3),\\ & (11, 4), (10, 5), (9, 6), (8,7) \\ 

    \end{tabular}
    \caption{Values of the first order term coefficient $f_{ab}$ in Eq.~\eqref{eqn:deffab} in the Walsh high-frequency expansion. All non-zero terms tabulated with magnitude greater than or equal to $2^{-5}$.}
    \label{table:fab}
\end{table}

We now apply this expansion to a few examples.
Consider a single spin $1/2$ driven by an up-down kick drive with 
\begin{equation}
    \hat{H}(t)= B_z \sigma^z+\Delta V_{\uparrow \downarrow,\text{n}}(t) \sigma^x.
\end{equation}
From the Baker-Campbell-Hausdorff expansion, we find that $H_\text{eff}=B_z \sigma^z+ B_z\Delta \sigma^y+\mathcal{O}(\Delta^2)$. The expansion is sensitive to the Floquet gauge in this case (see Sec.~\ref{sec:Expansion}), so using the non-symmetrized gauge with $f_{10}=1/4$ and coefficient $h_1=\frac{2}{T}\Delta \sigma^x$, we recover the same result. If we use the symmetrized gauge, the first order term vanishes such that $H_\text{eff}'=B_z \sigma^z +\mathcal{O}(\Delta^2)$ which is equivalent to a Schrieffer-Wolff transformation on $H_\text{eff}$ \cite{Bravyi2011}.

\begin{figure}[ht!]
    \centering
    \includegraphics[width=\linewidth]{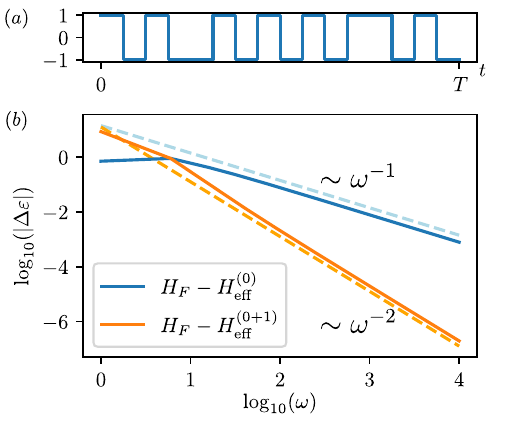}
    \caption{(a) Walsh function $W_{13}(t)$ with a sequency of 13. \\(b) difference between Walsh series and true Floquet Hamiltonian for $\hat{H}(t)=B_z \sigma^z+ W_2(t) B_x \sigma^x+W_{13}(t) B_y \sigma^y$. The eigenvalues, $\Delta \varepsilon$ of the remainder $H_F-H^{(0+1)}_\text{eff} \equiv H_F-H^{(0)}_\text{eff}-H^{(1)}_\text{eff}\sim \omega^{-2}$ demonstrating that the first order correction successfully canceled contributions $\mathcal{O}(\omega^{-1})$.
    }
    \label{fig:serieserror}
\end{figure}

To demonstrate the utility of the Walsh expansion, we study a two-tone driving of the form
\begin{equation}
    \hat{H}(t)=B_z \sigma^z+ W_2(t) B_x \sigma^x+W_{13}(t) B_y \sigma^y,
\end{equation}
where $W_a(t)$ is the Walsh function labelled with sequency. Due to the frequent flipping of $W_{13}(t)$ (Fig.~\ref{fig:serieserror}(a)), a BCH-type expansion involves several steps. However, from a single commutator in the Walsh expansion, we find the relevant first order correction to be
\begin{equation}
    H_{\text{eff}}^{(1)}=\frac{T}{16}B_x B_y\sigma^z.
    \label{eqn:complicateddrive}
\end{equation}

We validate this result by numerically calculating the full evolution, $H_F$, and studying the scaling with frequency $\omega=2\pi/T$ after subtracting off terms calculated from the series. Removing the zero order contribution, $H_\text{eff}^{(0)}\sim\mathcal{O}(\omega^0)$ (shown in blue in Fig.~\ref{fig:serieserror}(a)), means the dominant omega scaling is $\omega^{-1}$. Also removing the correct first order term, $H_\text{eff}^{(1)} \sim \mathcal{O}(\omega^{-1})$ (shown in orange in Fig.~\ref{fig:serieserror}(a)), gives a dominant scaling of $\omega^{-2}$. On a log-log plot, the power law dependence of the remaining dominant contribution in the series is a linear trend, and the slopes are in agreement with the predicted $\omega$ scaling demonstrates that Eq.~\eqref{eqn:complicateddrive} is correct.

While the Walsh basis effectively condenses the $N^2$ commutators in the BCH expansion to $N$ terms or less, there is a trade-off in the complexity of each individual commutator term. Here, symbolic approaches may be helpful in combination with the techniques developed in our work \cite{SandovalSantana2019}.

\pagebreak

\end{document}